\documentclass[twocolumn,epjc3]{svjour3} 

\usepackage{amssymb}
\usepackage{amsmath}
\usepackage{color}
\definecolor{darkgreen}{rgb}{0.05,0.85,0.05}

\usepackage{hyperref}
\hypersetup{colorlinks=true,
linkcolor=blue,
urlcolor=cyan,
citecolor=darkgreen}

\usepackage{cite}
\RequirePackage{fix-cm}
\RequirePackage{graphicx}
\journalname{Eur. Phys. J. C}

\begin{document}

\title{\boldmath 
Lifetimes of $b$-hadrons and mixing of neutral $B$-mesons: theoretical and experimental status 
}

\author{Johannes~Albrecht\thanksref{e1,addr1} 
        \and
        Florian~Bernlochner\thanksref{e2,addr2} 
        \and
        Alexander~Lenz\thanksref{e3,addr3}
        \and
        Aleksey~Rusov\thanksref{e4,addr3} 
}

\thankstext{e1}{e-mail: johannes.albrecht@tu-dortmund.de}
\thankstext{e2}{e-mail: florian.bernlochner@uni-bonn.de}
\thankstext{e3}{e-mail: alexander.lenz@uni-siegen.de}
\thankstext{e4}{e-mail: rusov@physik.uni-siegen.de}

\institute{TU Dortmund, Fakult\"at Physik, Otto-Hahn-Str. 4a, 44227 Dortmund, Germany \label{addr1}
  \and
Rheinische Friedrich-Wilhelms-Universität Bonn, Physikalisches Institut,
Nu\ss allee 12, 53115 Bonn, Germany 
\label{addr2}
  \and
Theoretische Teilchenphysik, Center for Particle Physics Siegen, Physik Department, Universität Siegen, \\ Walter-Flex-Strasse 3, 57068 Siegen, Germany  
\label{addr3}
}

\date{Received: -- / Accepted: --}

\maketitle
\begin{abstract}
In this article, we review the current status of $B$-mixing and $b$-hadron lifetimes both from experimental and theoretical points of view. 
Furthermore, we discuss the phenomenological potential of these observables for deepening our understanding of quantum chromodynamics (QCD) and for indirect searches for effects beyond the Standard Model (SM). 
In addition, we present new updated SM predictions for the mixing observables $\Delta M_{d,s}$, $\Delta \Gamma_{d,s}$ and $a_{fs}^{d,s}$. 
We conclude with an outlook on future prospects for theoretical and experimental improvements.
\\

\end{abstract}


\section{Introduction}
\label{intro}

\subsection{Colour meets Flavour}
\label{intro_CmF}
Lifetimes of hadrons are besides their mass the most fundamental
property of these compound objects. 
Their understanding is of principal importance for our grasp of the world of elementary particles.
Here, we will consider lifetimes of weakly decaying hadrons, 
that contain a heavy $b$-quark. In order to describe  these
quantities theoretically we need a quantitative control of the
hadronic binding effects both on the perturbative and the
non-perturbative side and of the underlying weak decay process.

Mixing of neutral mesons is a macroscopic quantum effect that is
heavily suppressed in the Standard Model. By now this effect is
experimentally well-established for kaons, $D^0$-, $B_d$- and
$B_s$-mesons. Again, the fundamental process triggering mixing is
governed by the weak interaction, which is partly obscured by
hadronic binding effects. Tests of the SM and searches for beyond SM (BSM) effects require a rigorous control of both perturbative and non-perturbative QCD contributions. 

\subsection{Lifetimes}
\label{intro_life}

The $b$-quark was discovered in 1977 at the E288 experiment at Fermilab in proton-nucleon collisions with a large mass
of around $m_b  \approx  4.25 \, {\rm GeV}$~\cite{E288:1977xhf}.
$B$-mesons were subsequently identified at PEP at SLAC and CESR at Cornell
\cite{CLEO:1980tem, Finocchiaro:1980gy, CLEO:1980oyr}, see Ref.~\cite{10.1063/1.55111} for an overview.

In similarity to the muon decay we expect the total decay rate of  weakly decaying $b$-quarks to be proportional to $m_b^5$, see Eq.~(\ref{eqn:muon}), and thus the lifetime to be very short e.g. in comparison to the lifetime of $D$-mesons. It thus came as a surprise when in 1983 the lifetime of $B$-mesons was measured to be of the order of 1.8\,ps by the MAC experiment at the PEP~\cite{Fernandez:1983az}\footnote{The MAC results were confirmed by the MarkII and DELCO experiments at PEP~\cite{Lockyer:1983ev,DELCO:1984jng} and the TASSO experiment \cite{TASSO:1984ult} at PETRA at the DESY laboratory in Hamburg.
}.
This was a first hint for the smallness of the CKM coupling of the $b$-quark to the $c$-quark, $V_{cb}$\footnote{This led L.~Wolfenstein to propose his famous parametrisation of the CKM matrix \cite{Cabibbo:1963yz,Kobayashi:1973fv} in the same year \cite{Wolfenstein:1983yz}.}.

Starting from the matrix element of a $b$-hadron $H_b$, decaying into a final state $X$ with certain quark quantum numbers and governed by the effective weak Hamiltonian ${\cal H}_{\rm eff}$ (see e.g. Ref.~\cite{Buchalla:1995vs} for a review), squaring it, performing a phase space integration, and finally summing over all possible final states $X$, we obtain the expression for the total decay rate of the $H_b$-hadron
\begin{eqnarray}
\Gamma ( H_b) = 
\sum \limits_{X} \int_{\rm PS} (2 \pi)^4 \delta^{(4)}
(p_{H_b} - p_X) 
\frac{|\langle X | {\cal H}_{eff} | H_b \rangle |^2 }
{2 m_{H_b}} 
\, .
\nonumber
\\
\label{total}
\end{eqnarray}
With the help of the optical theorem the total decay rate in Eq.~(\ref{total}) can be rewritten as
\begin{equation}
\Gamma(H_b) = \frac{1}{2 m_{H_b}} \langle H_b |{\cal T} | H_b \rangle \, ,
\end{equation}
with the transition operator
\begin{equation}
 {\cal T} = \mbox{Im} \; i \int d^4x \,
 T \left\{ {\cal H}_{eff}(x) {\cal H}_{eff} (0) \right\}\, ,
\label{eq:trans-oper}
\end{equation}
given by a discontinuity of a non-local double insertion of the effective $\Delta B = 1$ Hamiltonian ${\cal H_{\rm eff}}$. 
The transition operator in Eq.~\eqref{eq:trans-oper} can be further expanded in inverse powers of $b$-quark mass, which is with a value of $\sim 5$~GeV much larger than a typical hadronic scale of the order of a  few hundred MeV.
The resulting series in inverse powers of $m_b$ is referred to as heavy quark expansion (HQE).
First ideas for using of HQE in the theoretical treatment of heavy hadrons have started to {be} developed from 1973 onwards \cite{Nikolaev:1973uu} -- see e.g. the review \cite{Lenz:2014jha}. For a more detailed introduction and technical aspects of the heavy quark expansion, heavy quark symmetry and heavy quark effective theory (HQET), we refer to the review by Neubert~\cite{Neubert:1993mb}.

The main result of the HQE is that the total decay rate of the
bound state $H_b$ is given by the simple decay rate of a free $b$ quark, $\Gamma_b$, plus corrections depending on the decaying hadron $\delta \Gamma (H_b)$, which are suppressed by at least two powers of the $b$-quark mass $m_b$ relative to a hadronic scale $\Lambda$,
\begin{eqnarray}
\Gamma (H_b) = \frac{1}{\tau(H_b)} & = & 
\Gamma_b + \delta \Gamma (H_b),  
\nonumber \\
\delta \Gamma (H_b) & \propto & {\cal O}
\left( \frac{\Lambda^2}{m_b^2} \right),
\end{eqnarray}
with $\tau(H_b)$ being the lifetime of the hadron $H_b$.
The free $b$-quark decay has the same structure as the familiar muon decay
     \begin{eqnarray}
          \Gamma_b & = & \Gamma_0  \Bigr[
          N_c \left(|V_{ud}|^2 f(x_c,x_u,x_d) + 
                    |V_{cs}|^2 f(x_c,x_c,x_s) 
                    \right)
          \nonumber
          \\
           && 
          + \, f(x_c,x_e,x_{\nu_e}) +
           f(x_c,x_\mu,x_{\nu_\mu}) +
           f(x_c,x_\tau,x_{\nu_\tau}) 
           \nonumber
           \\
           && + \mbox{ CKM suppressed modes}
          \Bigl]
          \label{eq:free}
          \, ,
    \end{eqnarray}
with the number of colors $N_c$, phase space functions $f$ depending on mass ratios $x_q = m_q/m_b$, and the prefactor \begin{equation}
\Gamma_0 = \frac{G_F^2 m_b^5}{192 \pi^3} |V_{cb}|^2 \, ,
\label{eqn:muon}
\end{equation}
where $G_F$ denotes the Fermi constant.
The first line in Eq.~(\ref{eq:free}) describes the CKM leading non-leptonic decays $b \to c \bar{u} d$ and $b \to c \bar{c} s$, the second line CKM leading semi-leptonic decays $b \to c e \bar{\nu}_e $,  $b \to c \mu \bar{\nu}_\mu $ and $b \to c \tau \bar{\nu}_\tau $. 
The prefactor $\Gamma_0$ is strongly suppressed (thus leading to a long lifetime) by the small CKM element $V_{cb}$ and strongly enhanced by the large mass of the $b$-quark.
The dependence on $m_b^5$ is the source of large theory uncertainties in the prediction of the total decay rate.
{ However}, lifetime ratios are { theoretically} much cleaner, because there the free-quark decay rate, $\Gamma_b$, cancels completely,
\begin{eqnarray}
\frac{\tau(H_b)}{\tau(H'_b)}
& = & 1 + \left[ \delta \Gamma (H'_b) - \delta \Gamma (H_b) \right]
\cdot \tau(H_b)
\, .
\end{eqnarray}
Without knowing the size of higher-order QCD corrections, and with 
only very rough estimates for the non-perturbative matrix elements arising in the HQE, the naive expectation for lifetime ratios was in 1986~\cite{Shifman:1986mx} 
\begin{align}
& \left. \frac{\tau (B^+)}{\tau (B_d)} \right|^{\rm HQE \, 1986} \approx 1.1,
&
& \left. \frac{\tau (B_s)}{\tau (B_d)} \right|^{\rm HQE \, 1986} \approx 1
\, ,
\nonumber
\\
& \left. \frac{\tau (\Lambda_b)}{\tau (B_d)} \right|^{\rm HQE \, 1986} \approx 0.96
\, .
\label{eq:HQE_1986}
\end{align}
For the $B$-mesons this naive expectation was more or less confirmed experimentally. 

\begin{figure}[t]
\centering
\includegraphics[width=0.48\textwidth]{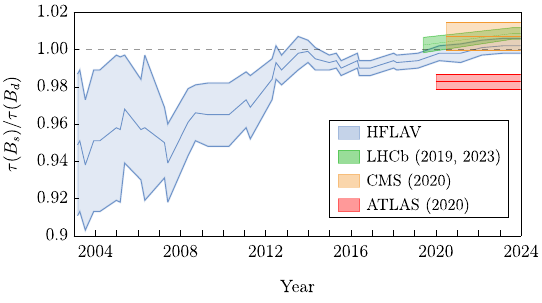}
\caption{History of the experimental averages of the lifetime of the $B_s$ meson, normalised to $\tau (B_d)$. We also indicated the most recent measurement by LHCb~\cite{LHCb-183, LHCb:2023sim}, CMS~\cite{CMS:2020efq} and ATLAS~\cite{ATLAS:2020lbz}, where the latter one seems to be in slight discrepancy with the average, see Section \ref{sec:Bs-Mix_Exp}.}
\label{fig:Bs}  
\end{figure}
\begin{figure*}[t]
\centering
\includegraphics[width=0.8\textwidth]{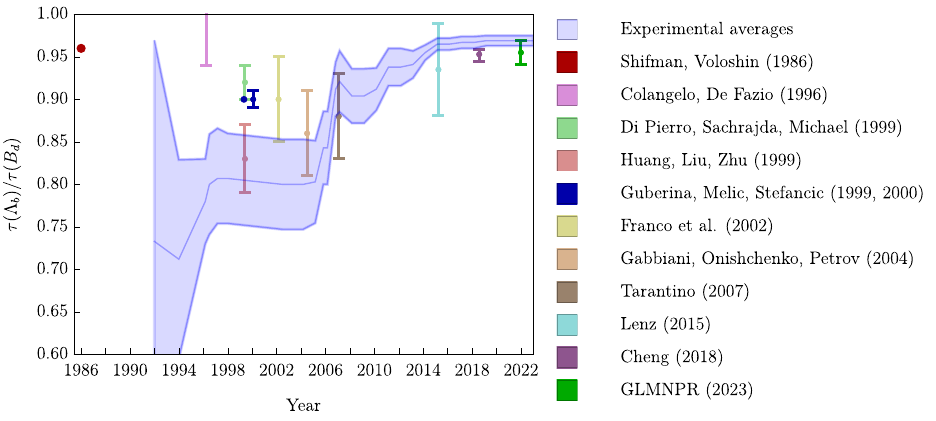}
\caption{Evolution of experimental extractions and theory determinations for the $\Lambda_b$ lifetime, updated from  Ref.~\cite{Gratrex:2023pfn}. GLMNPR (2023) refers to the study of Ref.~\cite{Gratrex:2023pfn}.}
\label{fig:tau-Lambda-b-to-tau-Bd-history-Exp-vs-Th}
\end{figure*}

Many experiments at the time have used the impact parameter of the tracks to deduce the $b$-hadron lifetime, a method that is largely independent of the boost of the $b$-hadron but extracts the average $b$-lifetime relying on Monte Carlo simulations, e.g. as used in Ref.~\cite{OPAL-RF1}. An alternative technique makes use of decays of the type $B\to J/\psi X$, which allows a very clean event selection, see Ref.~\cite{CDF-RF2} as an example. A third class of measurements uses fully reconstructed hadronic events, which reduces the use of simulations but requires a careful understanding of complex background structures, examples of this method include Refs.~\cite{ALEPH-RF3,DELPHI-RF4}. A historical overview of the results up to  1993 is given in Ref.~\cite{CERN-PPE-93-165}, the results before LEP accumulated at relatively low values of below one picosecond. LEP results in the early 1990's then stabilized and showed the results using $J/\psi$, hadronic or semileptonic events of around 1.5\,ps~\cite{CERN-PPE-93-165}.

The lifetime ratio $\tau (B^+)/\tau (B_d)$ has been measured first by the ARGUS and CLEO collaborations~\cite{CLEO-RF29,ARGUS-RF30,ARGUS-RF31} and was found to be close to unity, confirming the old theory estimates. The $B^0_s$ mesons were produced at LEP along with $B^+$ and $B^0$ but with limited statistics which required more advanced analysis techniques (see, for example, Refs.~\cite{OPAL-RF1,ALEPH-RF3,DELPHI-RF4}). These measurements indicated a $B_s$ lifetime in agreement with the lifetime of the $B_d$ mesons, albeit with significant uncertainties, see Ref.~\cite{CERN-PPE-93-165} for a discussion. 
{ A history of experimental measurements of the lifetime ratio $\tau (B_s)/\tau (B_d)$ is shown in Fig.~\ref{fig:Bs} 
(this is an updated and modified version of Fig.~1 from Ref.~\cite{Lenz:2022rbq}).}

The lifetime of the $\Lambda_b$ was found to be considerably lower than the lifetime of the $B_d$ meson. The $\Lambda_b$ was discovered at CERN in 1991 by the UA1 collaboration \cite{UA1:1991vse} and the first measurement of the $\Lambda_b$ lifetime was performed by the ALEPH collaboration using in 1992 \cite{ALEPH:1992yid} resulting in $\tau (\Lambda_b) = \left( 1.12_{-0.29}^{+0.32}  \pm 0.16 \right) $~ps. 
As can be read off Fig.~\ref{fig:tau-Lambda-b-to-tau-Bd-history-Exp-vs-Th}
(adapted from Ref.~\cite{Gratrex:2023pfn})
follow-up measurements confirmed the low values, which raised severe
doubts in the underlying assumption of quark hadron duality within the HQE, see e.g. Refs.~\cite{Shifman:2000jv,Bigi:2001ys}. 

Most measurements of $b$-hadron lifetimes rely on using the flight distance divided by the relativistic factors $\beta \gamma c$. To this end the velocity or momentum of the decaying $b$-hadron has to be either estimated or directly measured. 
The first measurements of the lifetime at LEP relied on the partial reconstruction of the $\Lambda_b$ via semileptonic decays and either the full reconstruction of a $\Lambda_c^+$ baryon or a more inclusive approach relying on the identification of a proton or $\Lambda$ baryon. The high momentum lepton of the semileptonic decay provides a clear trigger signature. The determination of $\tau (\Lambda_b)$ in this category of measurements relies on Monte Carlo simulations to estimate the momentum of $\Lambda_b$. However, these simulations inherently carry systematic uncertainties that are challenging to control. Examples include the modeling of the $b$-quark fragmentation distribution, $\Lambda_b$ polarization, and the dynamics and composition of semileptonic decay. The world average of all such measurements is $\tau (\Lambda_b) = 1.247^{+0.071}_{-0.069} \, \mathrm{ps} $ from Ref.~\cite{HeavyFlavorAveragingGroup:2022wzx}.
 
With the advent of large $b$-hadron samples from the Tevatron and the LHC new experimental techniques could be used to measure the $\Lambda_b$ lifetime: using fully hadronic decays allowed for the direct determination of the $\Lambda_b$ momentum and thus $\beta \gamma c$. The most precise determinations rely on using $\Lambda_b \to \Lambda_c \, \ell^+ \ell^-$ and the current world average is $\tau (\Lambda_b) = 1.471 \pm 0.009  \, \mathrm{ps} $~\cite{HeavyFlavorAveragingGroup:2022wzx}. The precise origin of the disagreement of these two averages is unknown, but a likely candidate seems an unaccounted systematic effect in the semileptonic measurements from the indirect determination of the $\Lambda_b$ momentum, or alternatively an unusual statistical fluctuation. 

A full summary of the current status of experimental measurements (taken from~\cite{HeavyFlavorAveragingGroup:2022wzx, Workman:2022ynf} which is based on Refs.~\cite{DELPHI:2003hqy,ALEPH:2000kte,ALEPH:1996geg,DELPHI:1995hxy,DELPHI:1995pkz,DELPHI:1996dkh,L3:1998pnf,OPAL:1995bfe,OPAL:1998msi,OPAL:2000,SLD:1997wak,CDF:1998pvs,CDF:2002ixx,CDF:2010ibe,D0:2008nly,D0:2014ycx,BaBar:2001mmd,BaBar:2002nat,BaBar:2002jxa,BaBar:2002war,BaBar:2005laz,Belle:2004hwe,ATLAS:2012cvl,CMS:2017ygm,LHCb:2014qsd,LHCb:2014bqh,CDF:2010gif,D0:2004ije,ALEPH:1997rqk,DELPHI:2000aij,OPAL:1997zgk,CDF:1998htf,DELPHI:2000gjz,OPAL:1997ufs,CDF:2011utg,LHCb:2013cca,LHCb:2014wet,LHCb:2017knt,CDF:1997axv,D0:2004jzq,LHCb:2021awg,CMS:2019bbr,ALEPH:2000cjd,LHCb:2012zwr,LHCb:2013dzm,CDF:2011kjt,D0:2016nbv,LHCb:2016crj,LHCb:2013odx,CDF:2012nqr,D0:2011ymu,ATLAS:2014nmm,ATLAS:2016pno,ATLAS:2020lbz,CMS:2015asi,CMS:2020efq,LHCb:2014iah,LHCb:2017hbp,LHCb:2016tuh,LHCb:2023sim,DELPHI:1995jet,OPAL:1995nmi,DELPHI:1996eqs,ALEPH:1996kuy,CDF:1996fsp,ALEPH:1997ake,DELPHI:1999con,DELPHI:2005zmk,D0:2007pfp,CDF:2009sse,D0:2012hfl,CMS:2013bcs,LHCb:2014wvs,CDF:2014mon,LHCb:2014wqn,LHCb:2014chk,LHCb:2014jst,LHCb:2016coe}) is given in Table \ref{tab:exp-data}.
\begin{table}[h]
\centering
\renewcommand{\arraystretch}{1.6}
    \begin{tabular}{|c||c|c|c|}
    \hline
         & $\tau \, [{\rm ps}]$ & $\Gamma \, [{\rm ps}^{-1}]$
         & $\tau (H_b)/\tau (B_d)$
         \\
         \hline
         \hline
   $B^+$ & $1.638(4) $ & $0.611(2) $ & $1.076(4) $
    \\
    \hline
   $B_d$ & $1.519(4)$  & $0.658(2)$ & $ 1 $
    \\
    \hline
    $B_s$ & $1.521(5)$ & $ 0.657(2) $  & $ 1.002(4)$
    \\
    \hline
    $B_c^+$ & $0.510(9)$ & $ 1.96(4) $  & $ 0.336(6)$
    \\
    \hline
    $\Lambda_b$ & $1.471(9)$ & $ 0.680(4) $  & $ 0.968(6)$
    \\
    \hline
    $\Xi_b^-$ & $1.572(40)$ & $ 0.636(16) $  & $ 1.035(27)$
    \\
    \hline
    $\Xi_b^0$ & $1.480(30)$ & $ 0.676(14) $  & $ 0.974(20)$
    \\
    \hline
    $\Omega_b^-$ & $1.64\left(^{+18}_{-17}\right)$ & $ 0.61\left(^{+7}_{-6}\right)$  & 
    $1.08\left(^{+12}_{-11}\right)$
    \\
    \hline
    \end{tabular}
    \caption{Status of the experimental determinations of the $b$-hadron lifetimes~\cite{HeavyFlavorAveragingGroup:2022wzx, Workman:2022ynf}}
    \label{tab:exp-data}
\end{table}
The current experimental values for $\tau(B^+)/ \tau (B_d)$, 
$\tau(B_s)/ \tau (B_d)$ and $\tau(\Lambda_b)/ \tau (B_d)$ from Table \ref{tab:exp-data} agree very well with the qualitative 
theory estimates from 1986. As will be described
below  there has been tremendous progress in theory both on the
perturbative and the non-perturbative aspects of the HQE, 
that allow now a quantitative estimate of the central values and uncertainties for lifetimes and lifetime ratios, see Section~\ref{sec:life_theory}. 
Currently, $b$-hadron lifetimes do not show any sign of violations of quark-hadron duality. 
Since the total decay rate is dominated by the leading quark-level decays like $b \to c \bar{c} s$, $b \to c \bar{u} d$ and $b \to c \ell \bar{\nu}$,
lifetimes are generally considered to be relatively insensitive to BSM effects. 
This attitude has, however, started to change
recently. BSM effects in CKM leading non-leptonic tree-level decays
are found to be not excluded by experiment, see e.g. Ref.~\cite{Lenz:2019lvd},
and they could lead to sizable effects in the
extraction of the CKM angle $\gamma$ \cite{Brod:2014bfa} or in the
decay rate difference of neutral $B_d$-mesons, $\Delta \Gamma_d$
\cite{Bobeth:2014rda}. 
Moreover, BSM effects in $b \to c \bar{c} s$ decays, could also lead
to modifications in the rare $b \to s \ell \ell$ transitions 
\cite{Crivellin:2023saq,Jager:2017gal,Jager:2019bgk}, where the
so-called $B$-anomalies are currently found experimentally, see e.g. Ref.~\cite{Albrecht:2021tul} for a review. 
Finally, recent measurements of hadronic two-body decays, 
like $\bar{B}_s^0 \to D_s^+ \pi^-$, differ significantly from the 
SM predictions~\cite{Fleischer:2010ca,Bordone:2020gao,Iguro:2020ndk,Cai:2021mlt} based on QCD factorisation (QCDf), a powerful framework developed for computations of non-leptonic decays in Refs.~\cite{Beneke:1999br,Beneke:2000ry}.
It is of course not unlikely that this discrepancy might root in an underestimation of power corrections within the QCDf approach, see e.g. the recent study of non-factorisable soft-gluon effects from light-cone sum rule in Ref.~\cite{Piscopo:2023opf}, but potential BSM effects in the underlying 
$b \to c \bar{u} d(s)$ transitions have also been investigated in e.g. Refs.~\cite{Iguro:2020ndk,Cai:2021mlt,Bordone:2021cca}
and an experimental verification of a CP violating BSM origin of this discrepancy was suggested in Ref.~\cite{Gershon:2021pnc}.
\subsection{Mixing}
\label{intro_mix}
\begin{figure}
\centering
 \includegraphics[width=0.48\textwidth]{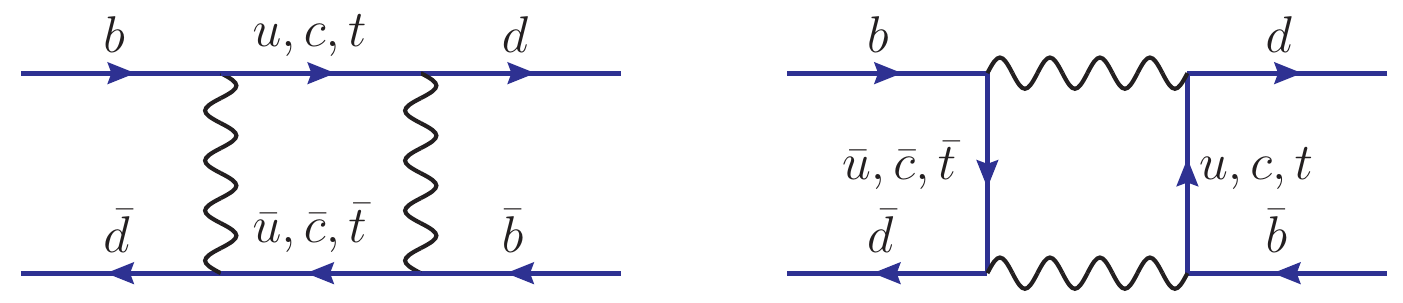}
\caption{Box diagrams describing the transition
of a $\bar{B}_d$ meson into a $B_d$ meson via the weak interaction.}
\label{fig:Box}  
\end{figure}

Box diagrams, see Fig.~\ref{fig:Box}, enable the transition of a $B_d$ meson -- defined by the quark content ($\bar{b}d$) -- into its anti-particle, $\bar{B}_d$.
This leads to the fact that mass eigenstates 
$B_{d, (H,L)}$ (with $H$= heavy and $L$ = light) of the neutral bottom-down mesons are actually defined as linear combinations of the quark eigenstates $B_d$ and~$\bar{B}_d$\footnote{We present here only the formulae for the $B_d$ system, the modification for the $B_s$ case is evident.}:
\begin{eqnarray}
B_{d, (H)} & = & p B_d - q \bar{B}_d \, ,
\nonumber
\\
B_{d, (L)} & = & p B_d + q \bar{B}_d \, .
\end{eqnarray}
Denoting the on-shell part of the box diagrams with
$\Gamma_{12}^d$ and the off-shell one with 
$M_{12}^d$, we can describe  three mixing observables (see
Refs.~\cite{Proceedings:2001rdi,Artuso:2015swg} for more details on the mixing formalism):
\begin{enumerate}
    \item Mass difference 
    \begin{eqnarray}
    \Delta M_d & = & M_{B_{d,(H)}} -  M_{B_{d,(L)}}
    \nonumber
    \\
    & = & 2 |M_{12}^d| + {\cal O} \left( 
    \left|\frac{\Gamma_{12}^d}{M_{12}^d} \right|^2
    \sin^2 \phi_{12}^d \right),
    \end{eqnarray}
    with $\phi_{12}^d = \arg (- M_{12}^d / \Gamma_{12}^d)$.
    
    \item Decay rate  difference 
    \begin{eqnarray}
    \Delta \Gamma_d & = & \Gamma_{B_{d,(L)}} -  \Gamma_{B_{d,(H)}}
    \nonumber
    \\
    & = & 
    2 |\Gamma_{12}^d| \cos \phi_{12}^d + {\cal O} \left(    \left|\frac{\Gamma_{12}^d}{M_{12}^d} \right|^2 \sin^2 \phi_{12}^d \right).
    \label{eq:def_DeltaGamma}
    \end{eqnarray}
    
    \item
    Flavour-specific CP asymmetries
    \begin{eqnarray}
    a_{\rm fs}^d & = &
    \left|  \frac{\Gamma_{12}^d}{M_{12}^d}
    \right|
     \sin \phi_{12}^d 
    + {\cal O} \left(\left| \frac{\Gamma_{12}^d}{M_{12}^d} \right|^2
    \sin^2 \phi_{12}^d \right).
    \end{eqnarray}   
    Since  flavour-specific CP asymmetries have so far only been measured in semi-leptonic decays they are also called 
    semi-leptonic CP asymmetries $a_{\rm sl}^{d}$.
\end{enumerate}
In the Standard Model, $|\Gamma_{12}^d / M_{12}^d| \sin \phi_{12}^d$ is tiny, of the order of  $-5 \cdot 10^{-4}$, in the $B_s$ system it is even smaller, of the order of $2 \cdot 10^{-5}$. Since mixing is described in the SM by suppressed loop diagrams, the corresponding observables are  generally expected to be very sensitive to BSM effects, see e.g. Refs.~\cite{Lenz:2010gu,Lenz:2012az}.
The mass and decay rate differences lead to time-dependent oscillations of the flavour eigenstates, which can be measured experimentally.

Mixing of neutral mesons was first discovered in the kaon system: Gell-Mann and Pais postulated in Ref.~\cite{Gell-Mann:1955ipe} in 1955 that neutral kaons are produced in two distinct flavors with a mismatch between the weak ($K_S / K_L$) and strong eigenstates ($K_0 / \overline K_0$). Consequent\-ly, the lifetimes and masses of the weak eigenstates may vary, and indeed, a long-lived neutral kaon was experimentally established in 1957 by Lande, Lederman, and Chinowsky using the Brookhaven Cosmotron experiment~\cite{Lande:1957bhp}. Pais and Piccioni outlined in 1955 in Ref.~\cite{Pais:1955sm} another consequence of the misalignment between weak and strong eigenstates: the occurrence of oscillations among the strong eigenstates over time. Ref.~\cite{Pais:1955sm} also outlined a proposal on how to establish the existence of mixing in an experiment: one may exploit that the strong interaction cross section of $\overline K^0$ is larger than $K^0$ at low energies to produce a regenerated $K_S$ beam. Such experiments were realized in 1961 using the Bevatron with the help of the Berkeley bubble chamber~\cite{Good:1961ik} and scintillation and Cherenkov counters~\cite{Fitch:1961zz}. In 1964, the observation of the decay of $K_L$ into two pions by Christenson, Cronin, Fitch, and Turlay would cement the violation of $CP$ in nature~\cite{Christenson:1964fg}, a discovery for which Cronin and Fitch were later awarded the Nobel prize. 

In 1987 the discovery of oscillations of neutral $B_d$ mesons  by the ARGUS experiment at DESY~\cite{ARGUS:1987xtv} came as a big surprise. At that time the mass of the top quark was expected to be of the order of 30 GeV, see e.g. Ref.\cite{Olsen:2023lrt}, while the observed strength of the mixing indicated a size of $m_t > 50 \, \mathrm{GeV}$~\cite{Schmidt-Parzefall:2007vsg}. This unexpected large size lead several authors to speculate about the existence of a heavy fourth generation $t'$ quark, see e.g. Ref.\cite{Tanimoto:1987wv}. The top quark was discovered at the Tevatron in 1995 by the CDF and D\O\ collaborations~\cite{CDF:1995wbb,D0:1995jca}. 

With the establishment of kaon and $B_d$ mixing, as well as the discovery of the top quark, the attention turned to the $B_s$ system. Besides much experimental effort until 2006 only a lower limit of $\Delta M_s > 14.5 \, \mathrm{ps}^{-1}$ at 95\% CL could be determined. D\O\ managed to improve this to a two-sided limit in Ref.~\cite{D0:2006oeb}, but the first measurement with a significance of 3 standard deviations was carried out by CDF~\cite{CDF:2006hcy}. To establish a difference in the decay rate, another collider was necessary: in 2012 the LHCb collaboration determined in Ref.~\cite{LHCb:2012via} for the first time a non-vanishing difference in the $B_s$ system.

The current experimental status of the mixing observables reads (taken from Ref.~\cite{HeavyFlavorAveragingGroup:2022wzx} which is based on the measurements 
of Refs.~\cite{ALEPH:1996ofb, L3:1998igq,OPAL:1996zny,CDF:1999jbn,DELPHI:2002aou,OPAL:1997pdn,OPAL:1997tfr,CDF:1998trv,CDF:1999jfn,CDF:1999hlp,D0:2006trs,BaBar:2002epc,BaBar:2001bcs,Belle:2002qxn,LHCb:2011tqa,LHCb:2012mhu,LHCb:2013fep,LHCb:2016gsk,OPAL:2000,BaBar:2002jxa,BaBar:2005laz,Belle:2004hwe,Belle:2002lms,hflav166,ATLAS-168,CDF-197,LHCb-198,LHCb-199,LHCb-200,LHCb-201,LHCb-183}):
\begin{align}
& \Delta M_d = 0.5065(19) \, \mbox{ps}^{-1}\!,
& & 
\Delta M_s = 17.765(6) \, \mbox{ps}^{-1}\!,
\label{eq:DeltaM-exp}
\\
& \frac{\Delta \Gamma_d}{\Gamma_d} = 0.001(10),
& & 
\Delta \Gamma_s = 0.083(5) \, \mbox{ps}^{-1}\!,
\label{eq:DeltaGamma-exp}
\\
& a_{\rm sl}^d = -21(14) \cdot 10^{-4},
& &
a_{\rm sl}^s = -60(280) \cdot 10^{-5}.
\label{eq:asl-exp}
\end{align}
The theory predictions for the mass differences require only the
knowledge of $M_{12}^{d,s}$, which is dominated by the top quark
contribution. The heavy top quark and the $W$-boson can be easily and
reliably integrated out and the theory precision is limited by our
knowledge of the non-perturbative matrix elements of $\Delta B = 2$
four-quark operators of mass dimension six. 
The experimental values for $\Delta M_d$ and $\Delta M_s$ are reproduced by the SM calculations, but with much higher uncertainties, see Eqs.~(\ref{eg:DeltaM-d-SM}) and (\ref{eg:DeltaM-s-SM}) below.

On the other hand, the theory determination of $\Gamma_{12}^{d,s}$ requires in addition a heavy quark expansion in the inverse powers of the $b$-quark mass, very similar to the one used for the $b$-hadron lifetimes.
Now we have to determine perturbative Wilson coefficients and 
non-perturba\-tive matrix elements of $\Delta B = 2$ operators of mass dimension six and higher.
The SM result for $\Delta \Gamma_s$ agrees well with experiment and has only a slightly worse accuracy, see Eq.~(\ref{eq:DeltaGamma-d-SM}) below, while $\Delta \Gamma_d$ is not yet measured with sufficient accuracy to establish it as different from zero, see Sec.~\ref{sec:mix_exp_Bd} for details. 

A BSM enhancement of the decay rate difference of $B_d$-mesons, $\Delta \Gamma_d$, might explain in part the long-standing $3.6 \sigma$ discrepancy of the D\O\ dimuon asymmetry \cite{Borissov:2013wwa,D0:2011hom,D0:2010hno,D0:2010sht}.
For the flavour-specific CP asymmetries we do not have a
non-zero measurement yet, and the current experimental bounds allow still for very sizable BSM effects. 
\section{Lifetimes}
\label{sec:Lifetimes}
\subsection{Theory}
\label{sec:life_theory}
\begin{figure*}[t]
\centering
\includegraphics[width=0.9\textwidth]{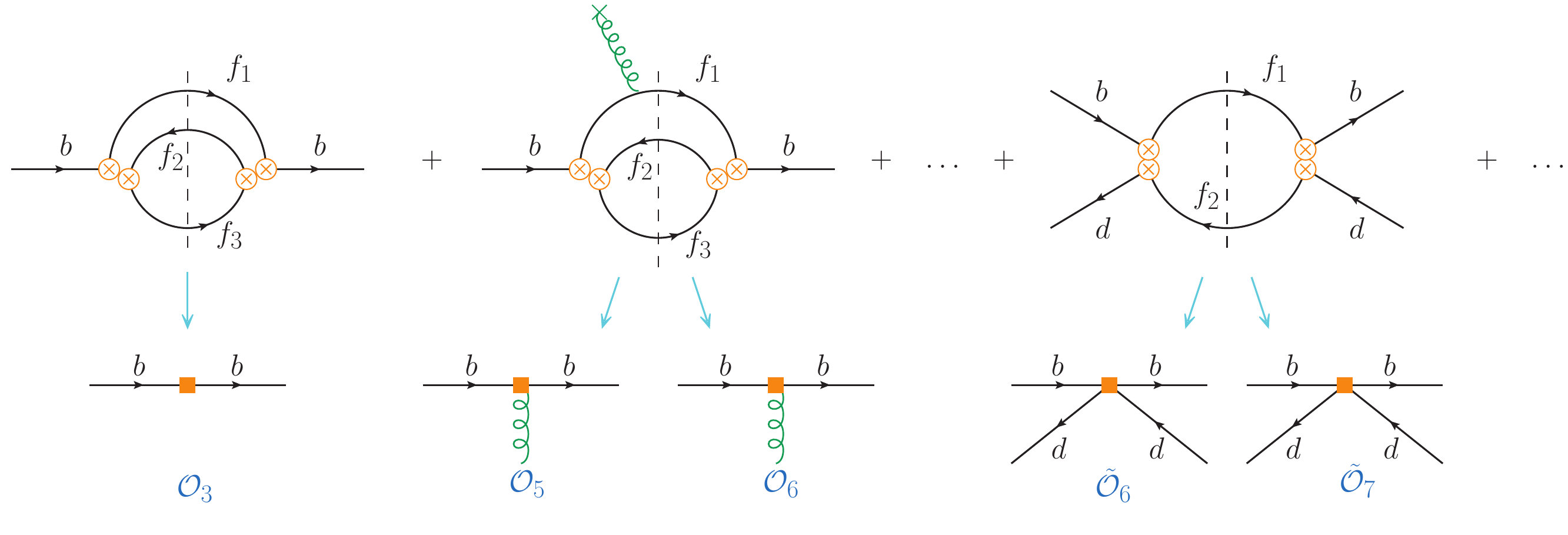}
\caption{Exemplary diagrams arising in the HQE of the total decay rate of the $B_d$-meson. The contribution on the left describes the free quark decay. The diagrams in the middle represent non-perturbative corrections to that, including contributions of the kinetic and the chromomagnetic operators at mass dimension five, of the Darwin operator at mass dimension six, and of operators of higher mass dimension. The diagrams on the right describe direct contributions of the spectator quark, encoded in four-quark operators of mass dimension six, seven and higher. $f_1, f_2, f_3 $ denote all possible combinations of fermions, into which the $b$-quark can decay.}
\label{fig:HQE-scheme}      
\end{figure*}
\begin{figure*}[t]
    \centering
    \includegraphics[width=0.25\textwidth]{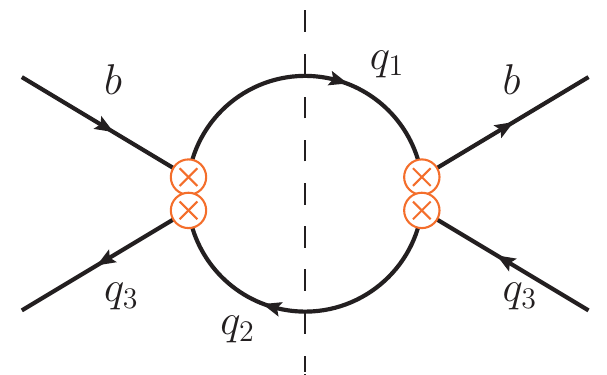}
    \qquad
    \includegraphics[width=0.25\textwidth]{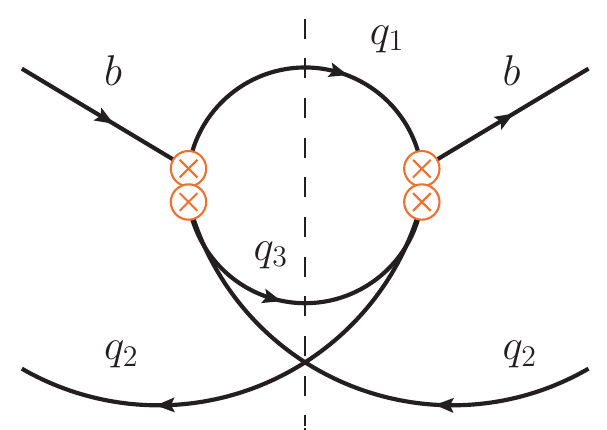}
    \qquad
    \includegraphics[width=0.28\textwidth]{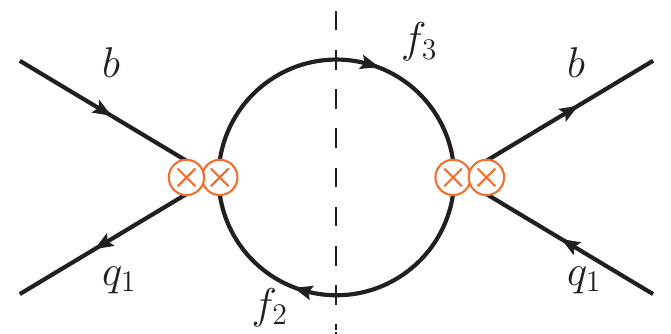}
    \caption{Diagrams describing the weak exchange (left), 
    Pauli interference (middle) and weak annihilation (right) topologies, at LO-QCD.  Here, $q_{1,2} = c, u$, \ $q_3 = d,s$, \ $f_1 = c, u, \ell$, and $f_2 = d, s, \nu_\ell$, with $\ell = e, \mu, \tau$.}
    \label{fig:WE-PI-WA-diagrams}
\end{figure*}

According to the HQE the total decay rate of a $b$-hadron $H_b$ can be expanded as (following the notation of Ref.~\cite{Lenz:2020oce})
\begin{eqnarray}
\Gamma(H_b) & = &  
\Gamma_3  +
\Gamma_5 \frac{\langle {\cal O}_5 \rangle}{m_b^2} + 
\Gamma_6 \frac{\langle {\cal O}_6 \rangle}{m_b^3} + ...  
\nonumber
\\
&&
+ 16 \pi^2 
\left(
\tilde{\Gamma}_6 \frac{\langle \tilde{\mathcal{O}}_6 \rangle}{m_b^3} 
+ \tilde{\Gamma}_7 \frac{\langle \tilde{\mathcal{O}}_7 \rangle}{m_b^4} + ... 
\right),
\label{eq:HQE}
\end{eqnarray}
where $\Gamma_i$ are short-distance functions, which can be computed perturbatively in QCD, i.e.
\begin{equation}
\Gamma_i = \Gamma_i^{(0)} + \frac{\alpha_s}{4 \pi} \Gamma_i^{(1)} 
+ \left(\frac{\alpha_s}{4 \pi}\right)^2 \Gamma_i^{(2)}+ \ldots\,, 
\label{eq:Gamma-i-pert-series}
\end{equation}
and $\langle {\cal O}_i \rangle \equiv
\langle H_b | {\cal O}_i |H_b  \rangle/(2 m_{H_b})$ denote non-pertur\-ba\-ti\-ve matrix elements of the corresponding $\Delta B = 0$ operators ${\cal O}_i$ of mass dimension $i$ in the effective theory. Note that at the same order in $1/m_b$ (starting from $1/m_b^3$), both two- and four-quark operator contributions can appear. The latter originate from loop-enhanced diagrams, as reflected by the explicit factor of $16 \pi^2$ in Eq.~\eqref{eq:HQE}, and to avoid confusion in the notation, we use a tilde to label them. 
The diagrammatic representation of Eq.~\eqref{eq:HQE} is shown in Fig.~\ref{fig:HQE-scheme}. 
Note that the perturbative Wilson coefficients of the two-quark contributions, ${\Gamma}_i$, are independent of the specific type of $b$-hadron $H_b$ considered, while the four-quark contributions $\tilde{\Gamma}_i$ and all non-perturbative matrix elements depend on it.

For discussing the theory state-of-the-art\footnote{The historic development of the theoretical status is discussed in more detail in Ref.~\cite{Lenz:2014jha}. A short, very recent theory summary including also results in the charm sector can be found in Ref.~\cite{Piscopo:2023jnu}.} we summarize first the available perturbative contributions to the semi-leptonic decays in the Table~\ref{tab:Coef-SL-summary}.  
For the free quark decay, $\Gamma_3$, QCD corrections up to order $\alpha_s^3$ (strong coupling)  are available, for the coefficients of the chromomagnetic operator, $\Gamma_5$, and the Darwin operator, $\Gamma_6$, QCD corrections up to order $\alpha_s^1$ are known, and for mass dimension seven and eight contributions the leading order QCD contributions have been determined.

\begin{table}[t]
\centering
\renewcommand{\arraystretch}{1.15}
\begin{tabular}{|l|l|l|}
\hline
$\Gamma_3^{(1)}$ & 1983 &
Ho-Kim, Pham~\cite{Hokim:1983yt}
\\
\hline
$\Gamma_3^{(2)}$ & 1997/98 & 
Czarnecki, Melnikov~\cite{Czarnecki:1997hc,Czarnecki:1998kt}
\\
 & 1999 &  
van Ritbergen~\cite{vanRitbergen:1999gs}
\\
& 2008 &
Melnikov~\cite{Melnikov:2008qs}
\\ 
& 2008 &  
Pak, Czarnecki~\cite{Pak:2008cp,Pak:2008qt}
\\ & 2008 &  
Dowling, Pak, Czarnecki~\cite{Dowling:2008ap}
\\ & 2008 & 
Bonciani, Ferroglia~\cite{Bonciani:2008wf}
\\
& 2009 &  
Biswas, Melnikov~\cite{Biswas:2009rb}
\\
& 2013 &  
Brucherseifer, Caola, Melnikov~\cite{Brucherseifer:2013cu} 
\\
&  2023 &
 Egner, Fael, Sch\"onwald \\
& &  Steinhauser~\cite{Egner:2023kxw}
\\
\hline
$\Gamma_3^{(3)}$ & 2020 &
Fael, Sch\"onwald, Steinhauser~\cite{Fael:2020tow}
\\
& 2021 & 
Czakon, Czarnecki, Dowling~\cite{Czakon:2021ybq} 
\\
&  2023 &
 Fael, Usovitsch~\cite{Fael:2023tcv} 
\\
\hline 
$\Gamma_5^{(0)}$ & 1992 &  
Bigi, Uraltsev, Vainshtein~\cite{Bigi:1992su}
\\
& 1992 &  Bigi, Blok, Shifman, 
\\
& &  Uraltsev, Vainshtein~\cite{Bigi:1992ne} 
\\
& 1992 & Blok, Shifman~\cite{Blok:1992hw,Blok:1992he}
\\
\hline
$\Gamma_5^{(1)}$ & 2013 &  
Alberti, Gambino, Nandi \cite{Alberti:2013kxa}  
\\
&  2014/15 &  
Mannel, Pivovarov, 
\\ 
& & Rosenthal~\cite{Mannel:2014xza,Mannel:2015jka}
\\
\hline 
$\Gamma_6^{(0)}$ & 1996 &  
Gremm, Kapustin~\cite{Gremm:1996df} 
\\
&  2017 &
 Mannel, Rusov, Shahriaran~\cite{Mannel:2017jfk}
\\
&  2022 & 
 Rahimi, Vos~\cite{Rahimi:2022vlv}
\\
\hline
$\Gamma_6^{(1)}$ & 2019  &  
Mannel, Pivovarov~\cite{Mannel:2019qel}
\\ 
& 2021 & 
Mannel, Moreno, Pivovarov~\cite{Mannel:2021zzr}
\\
& 2022 & Moreno~\cite{Moreno:2022goo}
\\
\hline 
$\Gamma_7^{(0)}$ & 2006 & 
Dassinger, Mannel, Turczyk~\cite{Dassinger:2006md}
\\
\hline
$\Gamma_8^{(0)}$ &  2010 &  
Mannel, Turczyk, Uraltsev~\cite{Mannel:2010wj}
\\
 & 2023 & 
Mannel, Milutin, Vos~\cite{Mannel:2023yqf}
\\
\hline
\end{tabular}
\caption{Summary of the theory status of the short-distance coefficients in the {\it semi-leptonic} decay widths.}
\label{tab:Coef-SL-summary}
\end{table}

Since for non-leptonic decays the calculations are technically more involved the expressions are typically only known to lower orders in $\alpha_s$ compared to the semi-leptonic case, see Table~\ref{tab:Coef-NL-summary}. 
Now the free quark decay is known to NLO accuracy and parts of the NNLO corrections have been estimated.
For the dimension-five chromo-magnetic operator very recently $\alpha_s$ corrections have been calculated for the massless charm-quark case.
The dimension-six Darwin operator contribution has only recently been determined to leading order in $\alpha_s$ and turned out to be sizable. 

Regarding the four-quark operator contribution (so-called spectator effects), there are three topologies contributing to the decay widths, 
named as the weak exchange (WE), Pauli interference (PI) and weak annihilation (WA), see Fig.~\ref{fig:WE-PI-WA-diagrams}\footnote{For the baryon case, the corresponding topologies are referred to as destructive Pauli interference, weak-exchange, and constructive Pauli interference, respectively.}. Up to now, the four-quark contributions  are known to orders $\alpha_s^1/m_b^3$ and $\alpha_s^0/m_b^4$, see Table~\ref{tab:Coef-NL-summary}.
All these perturbative terms contribute both to meson and baryon lifetimes. 

\begin{table}[th]
\renewcommand{\arraystretch}{1.15}
\centering
\begin{tabular}{|l|l|l|}
\hline
$\Gamma_3^{(1)}$ & 1983 & 
Ho-Kim, Pham~\cite{Hokim:1983yt}
\\ 
& 1991 & 
Altarelli, Petrarca~\cite{Altarelli:1991dx}
\\ 
& 1994 & 
Voloshin~\cite{Voloshin:1994sn}
\\
& 1994/95 &
Bagan, Ball, Braun/Fiol,
\\ 
&  & 
Goszinsky~\cite{Bagan:1994zd,Bagan:1995yf}
\\ 
& 1997/98 & 
Lenz, Nierste, Ostermaier \cite{Lenz:1997aa,Lenz:1998qp}
\\ 
& 2008 &         
Greub, Liniger~\cite{Greub:2000an,Greub:2000sy}
\\
& 2013 & 
Lenz, Krinner, Rauh \cite{Krinner:2013cja}
\\
\hline
$\Gamma_3^{(2)}$ & 2005 & 
Czarnecki, Slusarczyk, 
\\
& & Tkachov~\cite{Czarnecki:2005vr} ({\it  partly})
\\
\hline
$\Gamma_5^{(0)}$ & 1992 &  
Bigi, Uraltsev, Vainshtein~\cite{Bigi:1992su}
\\
&  1992 &  Bigi, Blok, Shifman, 
\\
& &  Uraltsev, Vainshtein~\cite{Bigi:1992ne} 
\\
& 1992 & Blok, Shifman~\cite{Blok:1992hw,Blok:1992he}
\\
\hline
$\Gamma_5^{(1)}$ & 2023 &  
Mannel, Moreno, Pivovarov~\cite{Mannel:2023zei}
\\
\hline
$\Gamma_6^{(0)}$ & 2020 &  
Lenz, Piscopo, Rusov~\cite{Lenz:2020oce}
\\
& 2020 & Mannel, Moreno, Pivovarov~\cite{Mannel:2020fts,Moreno:2020rmk}
\\
\hline
\hline
$\tilde{\Gamma}_6^{(0)}$ & 1979 
&  
Guberina, Nussinov, Peccei, 
\\
& & Ruckl~\cite{Guberina:1979xw} 
\\
& 1986 &
Shifman, Voloshin~\cite{Shifman:1986mx}
\\
& 1996 &  
Uraltsev~\cite{Uraltsev:1996ta}
\\
& 1996 & Neubert, Sachrajda~\cite{Neubert:1996we}
\\
\hline
$\tilde{\Gamma}_6^{(1)}$ & 2002 &  
Beneke, Buchalla, Greub, 
\\
& & Lenz, Nierste~\cite{Beneke:2002rj}
\\
& 2002 & 
Franco, Lubicz, Mescia, 
\\ 
& & Tarantino~\cite{Franco:2002fc}
\\
& 2013 &  
Lenz, Rauh~\cite{Lenz:2013aua}
\\
\hline
$\tilde{\Gamma}_7^{(0)}$ &  2003/04 &  
Gabbiani, Onishchenko,
\\
& & 
Petrov~\cite{Gabbiani:2003pq,Gabbiani:2004tp}
\\
\hline
\end{tabular}
\caption{Summary of the theory status of the short-distance coefficients in the {\it non-leptonic} decay widths.}
\label{tab:Coef-NL-summary}
\end{table}

Finally, for predicting lifetimes we need values for the arising non-perturbative matrix elements. Here we have to differentiate between meson and baryon cases. 
The matrix elements of two-quark operators for mesons are extracted from the fit of data on inclusive $B$-meson semi-leptonic decay width and moments of differential distributions. 
These fits are so far, however, only performed for the case of $B_d$- and $B^+$-mesons, while experimental input for $B_s$-mesons or $b$-baryons is still missing.
In addition, there exist several determination of the dimension-five two-quark operator matrix elements using the Lattice QCD and QCD sum rule methods, commonly with larger uncertainties. 

Matrix elements of the four-quark operators have so far only been determined with 3-loop HQET sum rules for $B_d$-, $B^+$- and $B_s$-mesons, as well as for $D$-mesons.
Lattice evaluations of these matrix elements are currently in progress, see Ref.~\cite{Black:2023vju}.
Values of matrix elements of colour-octet operators and so-called {\it eye-contractions}, determined using HQET sum rules, turned out to be very small -- here a lattice QCD confirmation of the numerical size would be very insightful.
For the matrix elements of four-quark operators of mass dimension seven we still can only use vacuum insertion approximation (VIA). 
Currently, we know much less about the matrix elements of baryons and we mostly have to rely on quark model estimates or on spectroscopic relations. A summary of the determinations of non-perturbative matrix elements, both for mesons and baryons, is presented in Table~\ref{tab:Lifetime-NP-input}.
\begin{table}[t]
\renewcommand{\arraystretch}{1.15}
\centering
\begin{tabular}{|l|l|l|}
\hline
$\langle Q_{5} \rangle_{B_d}$ 
& 1993/96 & 
QCD sum rule~\cite{Ball:1993xv,Neubert:1996wm} \\
& 2013-2023 &
Fit of inclusive data~\cite{Gambino:2013rza, Alberti:2014yda,Gambino:2016jkc,Bordone:2021oof,Bernlochner:2022ucr,Finauri:2023kte} 
\\
& 2017/18 &
Lattice QCD~\cite{Gambino:2017vkx,FermilabLattice:2018est}
\\
\hline
$\langle Q_{5} \rangle_{B_s}$ 
& 2011 & 
Spectroscopy relations~\cite{Bigi:2011gf}
\\
\hline
$\langle Q_{5} \rangle_{\cal B}$ 
& 2023 &
Spectroscopy relations~\cite{Gratrex:2023pfn} 
\\
\hline
$\langle Q_{6} \rangle_{B_d}$ 
& 1994/2022 &
EOM relation \cite{Mannel:1994kv, Lenz:2022rbq} 
\\
& 2013-2023 & 
Fit of inclusive data~\cite{Gambino:2013rza, Alberti:2014yda,Gambino:2016jkc,Bordone:2021oof,Bernlochner:2022ucr,Finauri:2023kte} 
\\
\hline
$\langle Q_{6} \rangle_{B_s}$ 
& 1994/2022 &
EOM relation~\cite{Mannel:1994kv, Lenz:2022rbq} 
\\
& 2011 &
Sum rule ~\cite{Bigi:2011gf}
\\
\hline
$\langle Q_{6} \rangle_{\cal B}$ 
& 2023 &
EOM relation~\cite{Gratrex:2023pfn} 
\\
\hline
\hline
$\langle \tilde{Q}_6 \rangle_{B_d}$ & 2017 &  
HQET sum rule~\cite{Kirk:2017juj}
\\
\hline
$\langle \tilde{Q}_6 \rangle_{B_s}$ & 2022  & 
HQET sum rule~\cite{King:2021jsq}
\\
\hline
$\langle \tilde{Q}_6 \rangle_{\Lambda_b}$ 
& 1996 & 
QCD sum rule~\cite{Colangelo:1996ta}
\\
\hline
$\langle \tilde{Q}_6 \rangle_{\cal B}$ 
& 2023 & 
NRCQM~\cite{Gratrex:2023pfn}
\\
\hline
$\langle \tilde{Q}_7 \rangle$ &  &  
VIA
\\
\hline
\end{tabular}
\caption{Status of determinations of the non-perturbative parameters for the $b$-hadron lifetimes. Here, $\cal B$ denotes the set of $b$-baryons $\{\Lambda_b, \Xi_b^0, \Xi_b^-, \Omega_b\}$.}
\label{tab:Lifetime-NP-input}
\end{table}

Based on this very advanced status of the theory framework we will present below theory predictions for total decay rates and  lifetime ratios.
\subsubsection{$B^+$- and $B_d$-mesons}
\begin{figure}[t]
\centering
\includegraphics[width=0.47\textwidth]{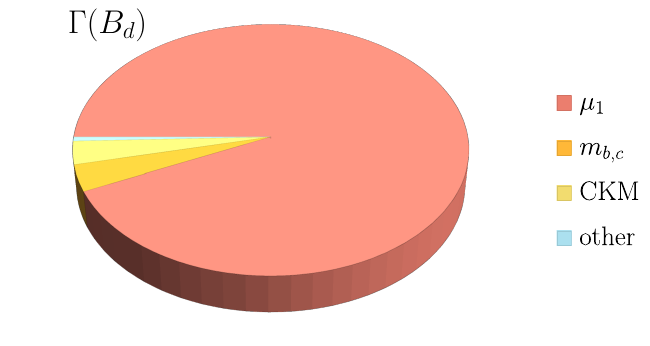}
\includegraphics[width=0.47\textwidth]{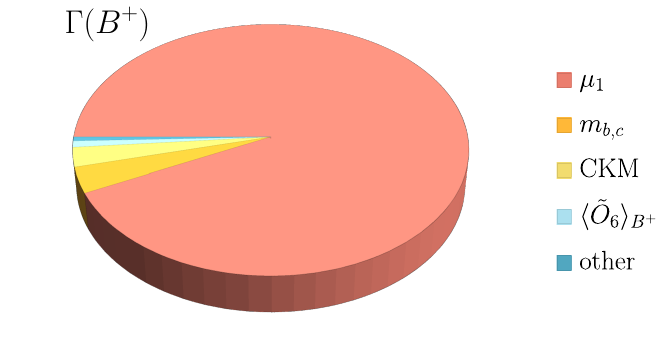}
\caption{Composition of the theoretical error in the HQE prediction of $\Gamma (B_d)$ and $\Gamma (B^+)$. In all pie-charts we show the relative size of the squared theoretical error, since we add all uncertainties in quadrature.}
\label{fig:Gamma_B_HQE}      
\end{figure}
\begin{figure}[t]
\centering
\includegraphics[width=0.47\textwidth]{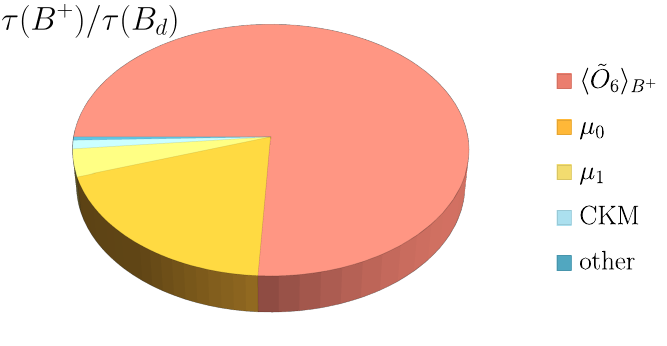}
\caption{Composition of the theoretical error in the HQE prediction of $\tau (B^+) / \tau (B_d)$.}
\label{fig:tau_B+Bd_HQE}      
\end{figure}

The total decay rate of the $B^+$- and $B_d$-mesons is predicted to be \cite{Lenz:2022rbq}
\begin{eqnarray}
\Gamma (B^+)  
& = &  
\left(0.58^{+ 0.11}_{-0.07}\right) \mbox{ps}^{-1},
\nonumber \\[1mm]
\Gamma (B_d)  & = &  
\left(0.63^{+ 0.11}_{-0.07}\right) \mbox{ps}^{-1}.
\end{eqnarray}
The small difference between the central values of the two decay rates stems mostly from the negative Pauli-interference term contributing to $\tilde{\Gamma}_6$ in the $B^+$-meson. The weak exchange contribution to the $B_d$-meson is numerically much smaller.
The composition of the large error (up to $19\%$ of the central value) is depicted in Fig.~\ref{fig:Gamma_B_HQE}. 
Since we add all uncertainties in quadrature, 
in all the pie-charts of this review we show the relative size of the {\it squared} theoretical errors. As one can see from Fig.~\ref{fig:Gamma_B_HQE},
the by far dominant uncertainty in $\Gamma (B^+)$ and $\Gamma (B_d)$ is given by the renormalisation scale dependence of the free-quark decay. Here, a first determination of $\alpha_s^2$ corrections to the free quark decay, i.e. $\Gamma_3^{(2)}$, including a proper choice of the quark mass scheme,  will improve the situation. Uncertainties due to CKM dependence and the value of quark masses are considerably smaller.

In the lifetime ratio $\tau(B^+)/ \tau(B_d)$ the free quark decay cancels, and due to isospin symmetry also all other two-quark contributions, i.e. $\Gamma_{5,6,...}$, vanish, 
and one finds therefore a much higher theory precision~\cite{Lenz:2022rbq}
\begin{eqnarray}
\frac{\tau (B^+)}{\tau (B_d)} 
& = & 
1.086 \pm 0.022 
 \, .
\end{eqnarray}
Now the theory error is only $2\%$!
The composition of this error is depicted in~Fig.~\ref{fig:tau_B+Bd_HQE}, 
and the dominant uncertainty stems from the size of the non-perturbative matrix elements of four-quark operators $\tilde {\cal O}_6$ followed by the renormalisation scale dependence of the Pauli interference term.
To reduce the former uncertainty first lattice evaluations of these matrix elements will be needed, see Ref.~\cite{Black:2023vju}, and to reduce
the second uncertainty NNLO-QCD corrections to the Wilson coefficient of the Pauli interference term, i.e. $\tilde{\Gamma}_6^{(2)}$, are required.
\subsubsection{$B_s$-mesons}
The theory prediction for the total decay rate of the
$B_s$-mesons \cite{Lenz:2022rbq} is
\begin{eqnarray}
\Gamma (B_s) & = & 
\left(0.63^{+ 0.11}_{-0.07}\right) \mbox{ps}^{-1}.
\end{eqnarray}
Up to the currently achievable theory precision we do not find any difference compared to the $B_d$-meson.
The composition of the theory error is more or less identical to the $B_d$-meson case, therefore we do not show it in a separate plot.

\begin{figure}[t]
\centering
\includegraphics[width=0.47\textwidth]{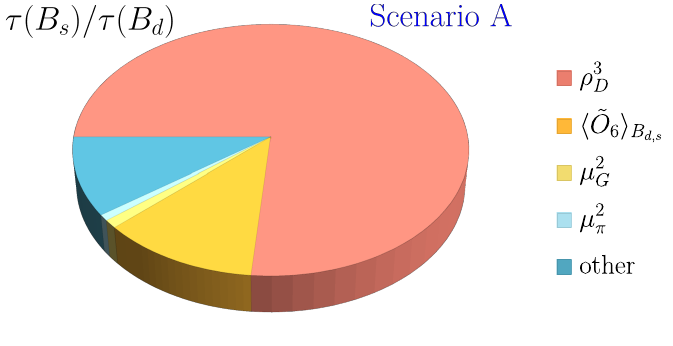}
\caption{Composition of the theoretical error in the HQE prediction of the lifetime ratio $\tau (B_s) / \tau (B_d)$.}
\label{fig:tau_BsBd_HQE}      
\end{figure}

Regarding the lifetime ratio $\tau (B_s)/\tau (B_d)$, now only the free quark decay cancels there, while two-quark contributions survive as $SU(3)_F$-symmetry breaking effects \cite{Lenz:2022rbq}.
In particular, we find a huge dependence of $\tau (B_s)/\tau (B_d)$ on the Darwin term $\Gamma_6 \, \langle {\cal O}_6 \rangle_{B_s}/m_b^3$. 
Here, the perturbative coefficient turned out to be very large \cite{Lenz:2020oce,Mannel:2020fts,Moreno:2020rmk,Piscopo:2021ogu}, and the size of the corresponding non-perturbative matrix elements is not directly known. For $B_d$- and $B^+$-mesons the matrix element $\langle {\cal O}_6 \rangle_{B_d}$ can be extracted from fits of data on inclusive semi-leptonic decay of the $B_d$- or $B^+$-meson~\cite{Bordone:2021oof,Bernlochner:2022ucr}. 
Unfortunately, these two extractions result in very different values, which was also not entirely settled in a subsequent study of Ref.~\cite{Finauri:2023kte}\footnote{The matrix element of the Darwin operator can also be determined from the matrix elements of four-quark operators, using equations of motion \cite{Lenz:2022rbq}. Depending on the chosen renormalisation scale one either reproduces the small value for the Darwin matrix element found by Ref.~\cite{Bernlochner:2022ucr} or the large value found by Ref.~\cite{Bordone:2021oof}.}.
To determine the ratio $\tau (B_s) /\tau  (B_d)$ we actually need the $SU(3)_F$-symmetry breaking combination $\langle {\cal O}_6 \rangle_{B_s} - \langle {\cal O}_6 \rangle_{B_d}$. The size of the $SU(3)_F$ breaking effects is currently not well understood. Therefore in Ref.~\cite{Lenz:2022rbq} two scenarios have been investigated:
\begin{itemize}
 \item Scenario A:
 large value of $\langle {\cal O}_6 \rangle_{B_d}$ from Ref.~\cite{Bordone:2021oof} and large $SU(3)_F$ breaking;
 \\
 
 \item Scenario B: 
 small value of $\langle {\cal O}_6 \rangle_{B_d}$ from Ref.~\cite{Bernlochner:2022ucr} and small $SU(3)_F$ breaking.
\end{itemize} 
These two scenarios lead to quite different predictions:
\begin{eqnarray}
\frac{\tau (B_s)}{\tau  (B_d) } 
& = & 
\left\{
\begin{array}{cc}
  1.028 \pm 0.011   &  \mbox{ Scenario A}
  \\[1.5mm]
   1.003 \pm 0.006 &  \mbox{ Scenario B}
\end{array}
 \right.
 \, .
\end{eqnarray}
Until the discrepant status of the size of the Darwin-term for the $B_d$-meson and the size of the corresponding $SU(3)_F$ breaking is clarified, we have to accept a very large spread of the HQE predictions, see Fig.~\ref{fig:ratio_B_HQEvsExp}.
The composition of the relative theory errors is shown in Fig.~\ref{fig:tau_BsBd_HQE}, with the dominant uncertainty originating from the Darwin operator contribution.
The next important uncertainty stems from the non-perturbative matrix elements of the four-quark operators~$\langle \tilde {\cal O}_6 \rangle$, both for the $B_d$- and $B_s$-meson.

\subsubsection{$b$-baryons}
The current theory predictions for the total decay rates of $b$ baryons~\cite{Gratrex:2023pfn} are
\begin{eqnarray}
\Gamma (\Lambda_b)  & = &  
(0.671^{+ 0.108}_{-0.071}) \mbox{ps}^{-1}, 
\nonumber \\[1mm]
\Gamma (\Omega_b^0) & = &
(0.591^{+ 0.108}_{-0.071}) \mbox{ps}^{-1},
\nonumber \\[1mm]
\Gamma (\Xi_b^0)  & = & 
(0.670^{+ 0.108}_{-0.071}) \mbox{ps}^{-1},
\nonumber \\[1mm]
\Gamma (\Xi_b^-)  & = &  
(0.622^{+ 0.104}_{-0.067}) \mbox{ps}^{-1}.
\label{eq:Gamma-b-baryons}
\end{eqnarray}
As in the meson case, the theory error for the total decay rate is completely dominated by the free quark decay. The small difference in the central values arises mostly due to the dimension-six four-quark operator contribution (described by $\tilde{\Gamma}_i$), given by peculiar combinations of the destructive Pauli interference, weak exchange, and constructive Pauli interference diagrams. 
The more pronounced spectator effect in the decay width of $\Omega_b^0$-baryon, rendering the latter smaller compared to other $b$-baryons, c.f.~Eq.~\eqref{eq:Gamma-b-baryons}, stems from the large value of the matrix element of the four-quark operator $\langle \tilde O_6 \rangle_{\Omega_b}$, see Ref.~\cite{Gratrex:2023pfn}.
\\
In lifetime ratios, the free-quark decay again cancels, and the current predictions read
\cite{Gratrex:2023pfn}
\begin{eqnarray}
\frac{\tau (\Lambda_b)}{\tau  (B_d) } 
 = 0.955 \pm 0.014,  
& \, &
\frac{\tau (\Omega_b)}{\tau  (B_d) } 
 = 1.081 \pm 0.042,
\nonumber \\
\frac{\tau (\Xi_b^0)}{\tau  (B_d) } 
 = 0.956 \pm 0.023,  
& \, & 
\frac{\tau (\Xi_b^0)}{\tau  (\Xi_b^-) } 
 = 0.929 \pm 0.028.
\end{eqnarray}
\begin{figure}
\centering
\includegraphics[width=0.47\textwidth]{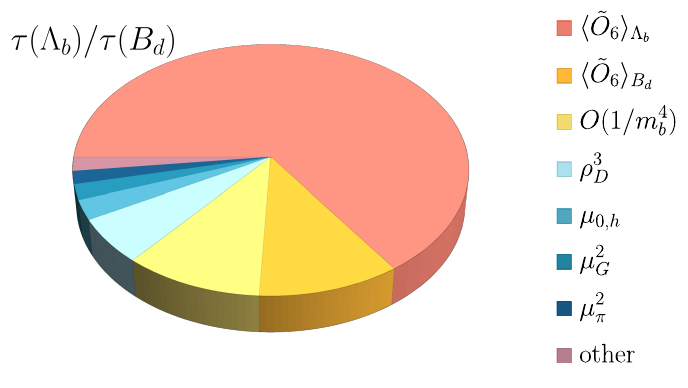}
\caption{Composition of the theoretical error in the HQE prediction of the lifetime ratio $\tau (\Lambda_b) / \tau (B_d)$.}
\label{fig:Gamma_Lambda_b_B_d_HQE}      
\end{figure}
\begin{figure}
\centering
\includegraphics[width=0.47\textwidth]{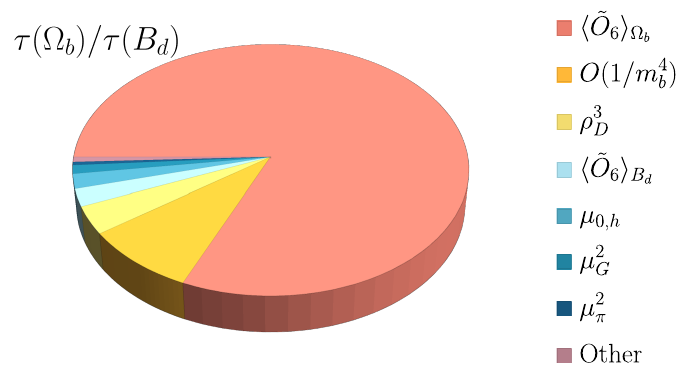}
 \caption{Composition of the theoretical error in the HQE prediction of 
 the lifetime ratio 
 $\tau (\Omega_b) / \tau (B_d)$.}
\label{fig:Gamma_Omega_b_B_d_HQE}      
\end{figure}
\begin{figure}
\centering
 \includegraphics[width=0.47\textwidth]{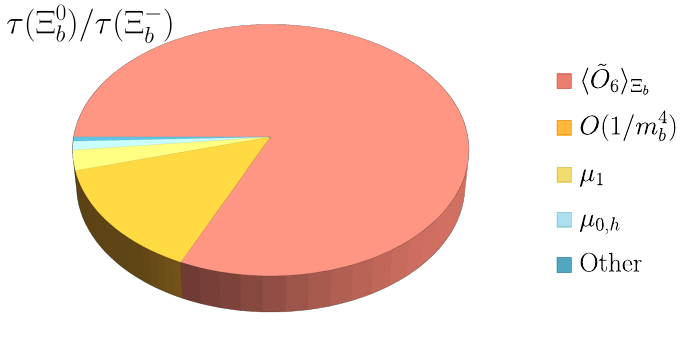}
 \caption{Composition of the theoretical error in the HQE prediction of 
 the lifetime ratio 
 $\tau (\Xi_b^0) / \tau (\Xi_b^-)$.}
\label{fig:Gamma_Xi_b_Xi_b_HQE} 
\end{figure}
\\
The relative theory error composition of
$\tau(\Lambda_b) / \tau (B_d)$,
$\tau(\Omega_b) / \tau (B_d)$
and
$\tau(\Xi_b^0) / \tau (\Xi_b^-)$
is presented respectively in Fig.
\ref{fig:Gamma_Lambda_b_B_d_HQE}, 
\ref{fig:Gamma_Omega_b_B_d_HQE} and  
\ref{fig:Gamma_Xi_b_Xi_b_HQE}  -- the one of
$\tau(\Xi_b^0) / \tau (B_d)$
looks very similar to
$\tau(\Lambda_b) / \tau (B_d)$ and thus we do not show it as a separate plot.
The by far dominant uncertainties stems from our limited knowledge of the values of the matrix elements of the four-quark operators.
In Ref.~\cite{Gratrex:2023pfn}, a common theoretical framework for these matrix elements was used, the non-relativistic constituent quark model (NRCQM)\footnote{An alternative common framework, the updated bag model, was used in  Ref.~\cite{Cheng:2023jpz}.}, see e.g. Ref.~{\cite{Guberina:1986gd}}. Therefore, further available results, like QCD sum rules estimate for the $\Lambda_b$-baryon \cite{Colangelo:1996ta,Zhao:2021lzd} were not taken into account. Here, clearly, future lattice QCD evaluations of these matrix elements would be highly desirable. Compared to the meson case, where the matrix elements of the four-quark operators are considerably more precisely known, in the baryon case the renormalisation scale dependence has much smaller relative uncertainty, albeit being of a similar overall size as for mesons.
\subsection{Experiment}
\label{sec:life_exp}
The LEP experiments performed the first measurements of $b$-baryon lifetimes, using partially reconstructed decays. With the large $b$-hadron samples available at the Tevatron and the LHC, the most precise measurements now come from fully reconstructed exclusive decays. A~complete and recent summary of the experimental status of $b$-baryon lifetimes is given in the recent HFLAV summary~\cite{HeavyFlavorAveragingGroup:2022wzx}, the 
world averages for the baryon lifetimes are summarized in Tab.~\ref{tab:exp-data}. 

$b$-baryons are copiously produced at high energy hadron colliders like the LHC. In principle, all three particle physics experiments have excellent capabilities to measure baryon lifetimes precisely, measurements of the $\Lambda_b$-lifetime have been performed at ATLAS~\cite{ATLAS-80}, CMS~\cite{CMS:2017ygm,CMS-132} and LHCb~\cite{LHCb:2014qsd,LHCb-133}. At the LHC, the lifetimes of $\Xi_b^-$, $\Xi_b^0$ and $\Omega_b^-$ have only been measured by the LHCb collaboration~\cite{LHCb-137,LHCb-138,LHCb-139,LHCb-140}. The currently ongoing Run~3 of the LHC gives excellent possibilities to improve these measurements, especially for LHCb which has re-optimized its detector to now run with a trigger-less readout and a full software trigger~\cite{LHCbUp}. This new setup allows high-statistics measurements, also of more difficult hadronic final states with minimal trigger biases. Especially for $\Omega_b^-$, but also for the two $\Xi_b^{0,-}$ states, the experimental uncertainty is partly larger than the theoretical one, hence more precise measurements, in particular of $\tau (\Omega_b)/ \tau (B_d)$ and $\tau (\Xi^0_b)/\tau (\Xi^-_b)$ would help to clarify the picture. 

{A new experimental possibility has been proposed to improve on the extraction of the Darwin operator matrix element and of corresponding $SU(3)_F$ breaking effects by measuring the moments} of the hadronic invariant mass distribution in semileptonic $B_s^0$-meson decays using a sum-of-exclusive technique~\cite{DeCian:2023ezb}. HQE parameters for the $B_s$- meson system can be extracted by a defined set of measurements that can be performed on current or soon to be collected data from LHCb on semi-leptonic $B_s^0$ decays. The interpretation of these data then requires an improved understanding of the excited $D_s^{**+}$ states, which can be gained from Belle~II and BES~III data. An interesting prospect would also be to extend this approach to study hadronic moments of inclusive semileptonic decays of $\Lambda_b^0$-baryons and other heavy hadrons. 

\subsection{Phenomenology}
\label{sec:life_pheno}
\begin{figure}[t]
\centering
\includegraphics[width=0.48\textwidth]{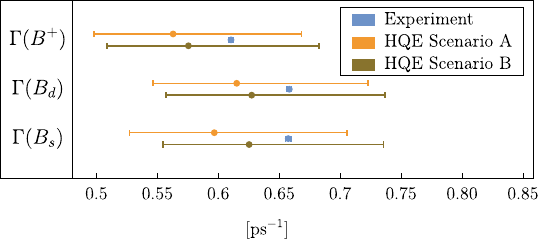}
\caption{HQE prediction of the total decay rates of the $B$-mesons compared to the experimental values.}
\label{fig:Gamma_B_HQEvsExp} 
\end{figure}
\begin{figure}[t]
\centering
\includegraphics[width=0.48\textwidth]{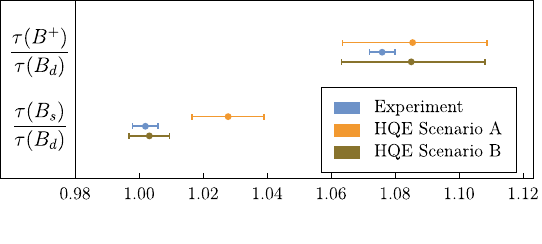}
\caption{HQE prediction of the lifetime ratios of the $B$-mesons compared to the experimental values.}
\label{fig:ratio_B_HQEvsExp}  
\end{figure}
Comparing experiment and theory~\cite{Lenz:2022rbq}, for the total decay rates of $B$-mesons we find a good agreement, albeit with much larger uncertainties in the HQE predictions, see Fig.~\ref{fig:Gamma_B_HQEvsExp}.
In the case of lifetime ratios, see Fig.~\ref{fig:ratio_B_HQEvsExp}, there is excellent agreement for $\tau(B^+)/\tau (B_d)$ with slightly larger theory uncertainties, while for the lifetime ratio $\tau(B_s)/ \tau (B_d)$ the situation is currently not settled, due to the unknown value of the size of the Darwin term and the corresponding $SU(3)_F$-symmetry breaking effects. In scenario~B, we find a perfect agreement of experiment and theory, in the case of scenario~A there is a tension arising at the level of 2 standard deviations.

\begin{figure}[t]
\centering
\includegraphics[width=0.48\textwidth]{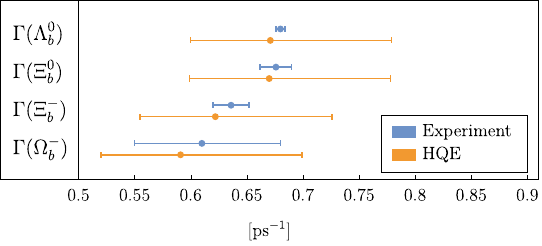}
\caption{HQE prediction of the total decay rates of $b$-baryons compared to the experimental values.}
\label{fig:Gamma_L_HQEvsExp} 
\end{figure}
\begin{figure}[t]
\centering
\includegraphics[width=0.48\textwidth]{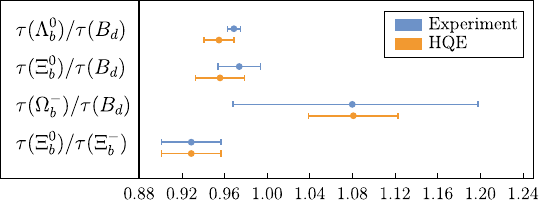}
\caption{HQE prediction of the lifetime ratios of $b$-baryons compared to the experimental values.}
\label{fig:ratio_L_HQEvsExp}   
\end{figure}

The most recent predictions for the total decay rates of $b$-baryons~\cite{Gratrex:2023pfn} agree also well with experiment. For the $\Xi_b$-baryons and in particular for the $\Lambda_b$-baryon the experimental uncertainties are considerably smaller than the theoretical ones, while in the case of $\Omega_b$-baryon experiment and theory have a similar size of uncertainties, see Fig.~\ref{fig:Gamma_L_HQEvsExp}.

For the lifetime ratio $\tau (\Lambda_b) /\tau (B_d)$ we find both in experiment and theory a small negative deviation from one~\cite{Gratrex:2023pfn}
\begin{eqnarray}
\frac{\tau(\Lambda_b)}{\tau (B_d^0)}
^{\rm HQE} & = & 1 - (0.045 \pm 0.014)
\, ,
\nonumber
\\
\frac{\tau(\Lambda_b)}{\tau (B_d^0)}
^{\rm Exp.} & = & 1 - (0.031 \pm 0.006)
\, .
\end{eqnarray}
Moreover, we confirm the experimentally observed lifetime splitting of the $\Xi_b^0$- and $\Xi_b^-$-baryons,
and coincidentally we find currently the same central value and uncertainty for the deviation from one~\cite{Gratrex:2023pfn}
\begin{eqnarray}
\frac{\tau(\Xi_b^0)}{\tau (\Xi_b^-)}
^{\rm HQE} & = & 1 - (0.071 \pm 0.028)
\, ,
\nonumber
\\
\frac{\tau(\Xi_b^0)}{\tau (\Xi_b^-)}
^{\rm Exp.} & = &  1 - (0.071 \pm 0.028)
\, .
\end{eqnarray}
For the $\Omega_b$-baryon we predict a larger lifetime compared to the $B^0_d$-meson, although here a clear experimental confirmation is still missing.
A comparison of the HQE predictions for lifetime ratios with the corresponding experimental data is presented in Fig.~\ref{fig:ratio_L_HQEvsExp}.

The excellent and precise agreement of theory and experiment for the lifetime ratio $\tau (B^+)/\tau (B_d)$ can be used to constrain BSM effects in non-leptonic tree-level decays~\cite{Lenz:2022pgw}, 
which will modify the Pauli-interference and weak exchange contributions, but again due to  iso\-spin symmetry the Darwin term will vanish. 
Such BSM effects in $b \to c \bar{u} d$ transitions could e.g. explain the discrepancy between predictions based on QCD factorisation and corresponding measurements of branching fractions of decays like $\bar{B}_s^0 \to D_s^+ \pi^-$~\cite{Fleischer:2010ca, Bordone:2020gao,Iguro:2020ndk,Cai:2021mlt}.
Ref.~\cite{Lenz:2022pgw} found that the lifetime ratio $\tau (B^+) / \tau (B_d)$ considerably shrinks the allowed parameter space in Ref.~\cite{Cai:2021mlt}, needed to explain the $\bar{B}_s^0 \to D_s^+ \pi^-$ anomalies by BSM effects.
Moreover, future, more precise measurement of the semi-leptonic CP asymmetry $a_{sl}^d$ have the potential to exclude or confirm a BSM contribution in the  $b \to c \bar{u} d$ channel.

BSM effects in $b \to c \bar{c}s $ decays could be the origin of the $B$-anomalies, and they will modify the lifetime ratio $\tau (B_s) / \tau (B_d)$ \cite{Lenz:2019lvd,Jager:2017gal,Jager:2019bgk}.
The same lifetime ratio will be modified by lepto-quark models explaining the $B$-anomalies and introducing large $b \to (d,s) \tau \tau$ transitions
\cite{Bobeth:2014rda,Bobeth:2011st,Bordone:2023ybl}. For a proper investigation of the BSM effects in $\tau (B_s) / \tau (B_d)$ the modification of the Darwin contribution due to BSM effects has to be taken into account, which is so far missing in the literature.

\subsection{Outlook to the Charm system}
\label{sec:life_charm}
The applicability of the HQE in the charm system suffers from lower values
of the charm quark mass and from larger values of the strong coupling. 
On the other hand, this offers an enhanced sensitivity  to higher orders in $1/m_c$ and $\alpha_s (m_c)$. 
Recent studies of charmed meson~\cite{King:2021xqp,Gratrex:2022xpm} and charmed baryon lifetimes~\cite{Gratrex:2022xpm} show within larger theory uncertainties an agreement with measurements, which increases considerably our confidence in the applicability of the HQE for the $b$-system.
\section{Mixing}
\label{sec:Mixing}
\subsection{Theory}
\label{sec:mix_theory}

\subsubsection{Mass difference}
The calculation of the dispersive part of the box-diagram depicted in Fig.~\ref{fig:Box}, $M_{12}^q$ (with $q=d,s$), yields
\begin{eqnarray}
M_{12}^q & = & \frac{G_F^2}{12 \pi^2}
\lambda_{tq}^2 M_W^2 S_0(x_t) \hat{\eta}_B
f_{B_q}^2  M_{B_q} B_1^q \, ,
\label{M12}
\end{eqnarray}
with the masses of the $W$-boson, $M_W$, and of the $B_q$-meson, $M_{B_q}$, and the CKM structure $\lambda_{tq} := V_{tq}^* V_{tb}$.
The perturbative 1-loop calculation is encoded in the Inami-Lim function
$ S_0(x_t)$ \cite{Inami:1980fz},
with $x_t= \bar{m}_t^2(\bar{m}_t)/M_W^2$, where $\bar{m}_t(\bar{m}_t)$ is
the $\overline{MS}$-mass \cite{Bardeen:1978yd} of the top quark.
Perturbative 2-loop QCD corrections to $M_{12}^q$ have been determined in 1990 by Buras, Jamin and Weisz and are contained in the factor
$\hat{\eta }_B \approx 0.84$
\cite{Buras:1990fn}.
Hadronic effects that describe the binding of the quarks into meson states
are encoded in the matrix element of the arising
$\Delta B=2$ four-quark operator
\begin{eqnarray}
\tilde{Q}_{6,1}^q & = & (\bar q^\alpha \gamma_\mu (1- \gamma_5) b^\alpha)
(\bar q^\beta  \gamma^\mu (1- \gamma_5) b^\beta), 
\label{Q}
\end{eqnarray}
with the colour indices $\alpha$ and $\beta$  of the $b$- and $q$-quark spinors. For historic reasons the hadronic matrix
element\footnote{Throughout this review we will
use the conventional relativistic normalisation for
the $B_q$ meson states, i.e. $\langle B_q | B_q \rangle = 2 E V$
($E$: energy, $V$: volume).} of this operator is parametrised in terms of a decay constant $f_{B_q}$ and a renormalisation scale dependent bag parameter $B_1^q$:
\begin{eqnarray}
\langle \tilde{Q}_{6,1}^q  \rangle & \equiv & 
\langle \bar{B_q}| \tilde{Q}_{6,1}^q  |B_q \rangle 
= \frac{8}{3}  M_{B_q}^2  f_{B_q}^2 B^q_1 (\mu).
\end{eqnarray}
In vacuum insertion approximation the
bag parameter $B^q_1$ receives the value $1$.\footnote{Sometimes a different notation for the QCD-corrections and the bag parameter is used in the literature
(e.g. by the Flavour Lattice Averaging Group (FLAG): \cite{FlavourLatticeAveragingGroupFLAG:2021npn}),
$(\eta_B, \hat{B}_1^q)$ instead of $(\hat{\eta}_B, B_1^q)$ with
\begin{eqnarray}
\hat{\eta}_B B_1^q & =: & 
\eta_B \hat{B}_1^q = 
\eta_B \, \alpha_s(\mu)^{-\frac{6}{23}} 
\left[1 + \frac{\alpha_s(\mu)}{4 \pi} \frac{5165}{3174} \right] B_1^q\; .
\end{eqnarray}
The parameter $\hat{B}_1^q$ has the advantage
of being renormalisation scale and scheme independent.}
In the SM, only the operator $\tilde{Q}_{6,1}^q$ in Eq.~(\ref{Q}) contributes to the mass difference of neutral $B_q$ mesons, while in extensions of the SM additional operators, denoted by $\tilde{Q}_{6,2}^q,...,\tilde{Q}_{6,5}^q$, with new Dirac and new colour structures can arise. 

The non-perturbative matrix elements have either been determined by lattice QCD simulations or by sum rule evaluations in HQET that require the calculation of discontinuities of three-loop diagrams. 
In the $B_s$-meson case mass effects of the strange quark had to be taken into account in the sum rule calculation. First steps towards a determination of the four-loop contribution to the sum rules have been undertaken in Refs.~\cite{Grozin:2017uto,Grozin:2018wtg}.
Interestingly, it turned out, that the elaborate determinations of the bag parameters for mixing in Refs.~\cite{Kirk:2017juj,FermilabLattice:2016ipl,Dowdall:2019bea,Grozin:2016uqy,King:2019lal} yield results that are very close to vacuum insertion approximation. 
In contrast to the lifetime case, where $\Delta B=0$ four-quark operators can also form eye-contractions, this option is excluded in the mixing case and it makes their non-perturbative determination simpler.
A summary of determinations of the non-perturbative matrix elements for mixing is presented in Table~\ref{tab:mixing-NP-input}.

\begin{table}[t]
\renewcommand{\arraystretch}{1.15}
\centering
\begin{tabular}{|l|l|l|}
\hline
\multicolumn{3}{|c|}{Lattice QCD} \\
\hline
$\langle \tilde{Q}_{6,1-5} \rangle_{B_{d,s}}$ & 2016 
& FNAL-MILC~\cite{FermilabLattice:2016ipl}
\\
\hline
$\langle \tilde{Q}_{6,1-5} \rangle_{B_{d,s}}$ & 2019 
& HPQCD~\cite{Dowdall:2019bea}
\\
\hline 
\hline
\multicolumn{3}{|c|}{HQET sum rule} \\
\hline
$\langle \tilde{Q}_{6,1} \rangle_{B_d}$ & 2017 &  
Grozin, Klein, Mannel, 
\\ 
& & Pivovarov~\cite{Grozin:2016uqy}
\\
\hline
$\langle \tilde{Q}_{6,1-5} \rangle_{B_d}$ & 2017 &  
Kirk, Lenz, Rauh~\cite{Kirk:2017juj}
\\
\hline
$\langle \tilde{Q}_{6,1-5} \rangle_{B_s}$ &2019 & 
King, Lenz, Rauh~\cite{King:2019lal}
\\
\hline
\end{tabular}
\caption{Status of determinations of the non-perturbative parameters for $B_q$-mixing.}
\label{tab:mixing-NP-input}
\end{table} 

We update the SM prediction for the mass difference from Refs.~\cite{Lenz:2019lvd,DiLuzio:2019jyq} with new CKM input and a new combination of lattice QCD and HQET sum rule results for the matrix elements of the four quark operators including correlations~\cite{Greljo:2022jac}
\begin{eqnarray}
\Delta M_d & = & 
 (0.535 \pm 0.021) \, \mbox{ps}^{-1},
\label{eg:DeltaM-d-SM}
\\
\Delta M_s & = & 
 (18.23 \pm 0.63) \, \mbox{ps}^{-1}.
\label{eg:DeltaM-s-SM}
\end{eqnarray}

\begin{figure}[t]
\includegraphics[width=0.5\textwidth]{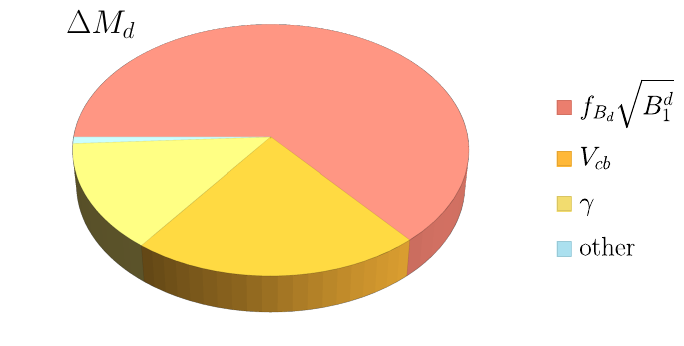}
\caption{Composition of the theoretical error in the SM prediction of 
$\Delta M_d$.}
\label{fig:Delta_M_d}      
\end{figure}
\begin{figure}[t]
  \includegraphics[width=0.5\textwidth]{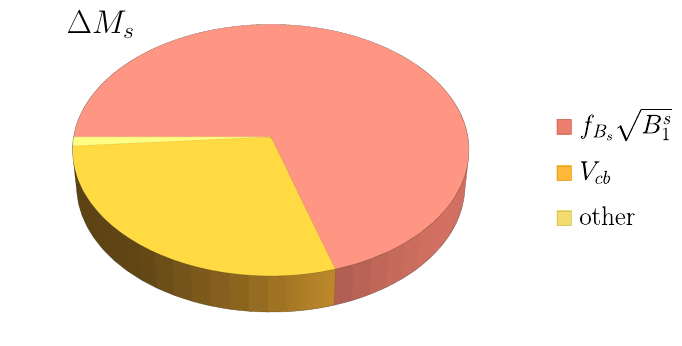}
\caption{Composition of the theoretical error in the SM prediction of 
$\Delta M_s$.}
\label{fig:Delta_M_s}      
\end{figure}

\noindent
The composition of the theoretical error is depicted in Fig.~\ref{fig:Delta_M_d} and Fig.~\ref{fig:Delta_M_s} and the uncertainties are dominated by our limited knowledge of the size of the matrix element of the operator $\tilde{Q}_{6,1}^q$ and the precision in the CKM element $V_{cb}$ -- for the mass difference $\Delta M_d$ the CKM angle $\gamma$ gives also a sizable contribution.
\subsubsection{Decay rate difference and flavour specific CP asymmetries}
\begin{figure}
\centering
 \includegraphics[width=0.30\textwidth]{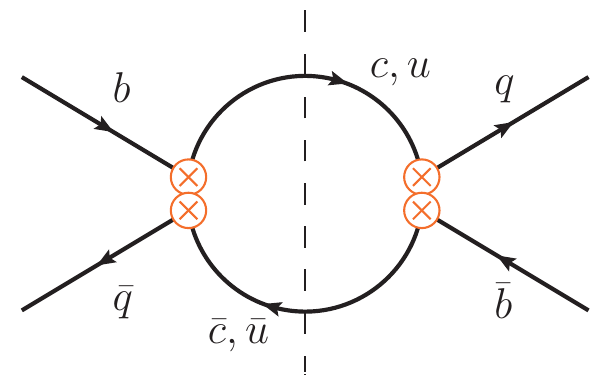}
\caption{Leading one-loop diagram contributing to $\Gamma_{12}^q$.}
\label{fig:Gamma12s}      
\end{figure}

In the calculation of the absorptive part of the box-diagram depicted in Fig.~\ref{fig:Box}, 
$\Gamma_{12}^q$, one starts again with integrating out the heavy $W$-boson to obtain the effective weak $\Delta B=1$
Hamiltonian, which is then used to determine mixing diagrams like
the one shown in Fig.~\ref{fig:Gamma12s}. 
Expanding in inverse powers of the $b$-quark mass one gets 
\begin{eqnarray}
\Gamma_{12}^q
 & = &  
 16 \pi^2 
\left( 
  \tilde{\Gamma}_6^q \frac{\langle \tilde{\mathcal{O}}_6 \rangle_{B_q}}{m_b^3} 
+ \tilde{\Gamma}_7^q \frac{\langle \tilde{\mathcal{O}}_7 \rangle_{B_q}}{m_b^4} + ... 
\right),
\label{eq:HQEmix}
\end{eqnarray}
which looks very similar to the four-quark contribution in the HQE of the total decay rate in Eq.~(\ref{eq:HQE}).
In the lifetime diagram, depicted in Fig.~\ref{fig:HQE-scheme}, there was an incoming $b$-quark and an outgoing $b$-quark, while in the mixing diagrams shown in Fig.~\ref{fig:Gamma12s} one has an incoming $b$-quark and an outgoing anti-$b$-quark.
The operators in Eq.~(\ref{eq:HQEmix}) are four-quark $\Delta B=2$ operators, while they were $\Delta B=0$ operators in the lifetime case.

In the SM there are three operators contributing to $\Gamma_{12}^q$ at mass dimension-six: $\tilde{Q}_{6,1}^q$, which contributes also to $M_{12}^q$ in the SM and  $\tilde{Q}_{6,2}^q$ and $\tilde{Q}_{6,3}^q$ which contribute to $M_{12}^q$ only in extensions of the SM. At leading order in $1/m_b$ and $\alpha_s$ these three operators are linear dependent, see e.g. Ref.~\cite{Beneke:1996gn}.
At mass dimension-seven there are either operators of the form
 $m_q \tilde{Q}_{6,4}^q$
 and
 $m_q \tilde{Q}_{6,5}^q$
 ($q=d,s$)
 arising, or 
genuine new operators, that consist of four-quark fields and one covariant derivative.
Thus for the SM determination of $\Gamma_{12}^q$ the full BSM basis
of mass dimension-six operators determined in Refs.~\cite{Kirk:2017juj,FermilabLattice:2016ipl,Dowdall:2019bea,King:2019lal} is needed and in addition a determination of the new genuine mass dimension-seven operators. For a long time phenomenological applications had to rely on VIA for the dimension-seven operators, but recently a first determination on the lattice was performed by the HPQCD collaboration~\cite{Davies:2019gnp}.
This pioneering work confirmed the expectation of the  VIA assumption also for dimension-seven operators. On the other hand, the results of Ref.~\cite{Davies:2019gnp} show very large uncertainties since ${\cal O} (\alpha_s)$ corrections to the  matrix elements of the dimension-seven operators have not yet been included.
Here additional studies, either on the lattice or with HQET sum rules are highly desirable.

\begin{table}[t]
\renewcommand{\arraystretch}{1.15}
\centering
\begin{tabular}{|l|l|l|}
\hline
$\tilde{\Gamma}_6^{(1)}$ & 1998 &
Beneke, Buchalla, Greub, 
\\ 
&& Lenz, Nierste~\cite{Beneke:1998sy}
\\
& 2003 &
Beneke, Buchalla, Lenz, Nierste  \cite{Beneke:2003az}
\\
& 2003 &   
Franco, Lubicz, Mescia, Tarantino~\cite{Ciuchini:2003ww}
\\
& 2006 &
Lenz, Nierste~\cite{Lenz:2006hd}
\\
\hline
$\tilde{\Gamma}_6^{(2)}$ & 2017 &
Asatrian, Hovhannisyan, Nierste,
\\
& &
Yeghiazaryan~\cite{Asatrian:2017qaz}
\\
& 2020 &
Asatrian, Asatryan, Hovhannisyan,
\\
& & 
Nierste, Tumasyan, Yeghiazaryan~\cite{Asatrian:2020zxa}
\\
& 2021 &
Gerlach, Nierste, Shtabovenko,
\\
& & 
Steinhauser~\cite{Gerlach:2021xtb,Gerlach:2022wgb,Gerlach:2022hoj}
\\
\hline
$\tilde{\Gamma}_7^{(0)}$ & 1996 &  
Beneke, Buchalla~\cite{Beneke:1996gn}
\\
\hline
$\tilde{\Gamma}_8^{(0)}$ & 2007 &  
Badin, Gabbiani, Petrov~\cite{Badin:2007bv}
\\
\hline
\end{tabular}
\caption{ Summary of theory status of short-distance contributions in $\Gamma_{12}^q$.}
\label{tab:Coef-Gamma12-summary}
\end{table}

On the perturbative side the theory status of $\Gamma_{12}^q$ is very advanced, see Table \ref{tab:Coef-Gamma12-summary}, and recently also NNLO-QCD corrections to $\tilde{\Gamma}_6^q$ became available including charm quark mass effects up to order $(m_c/m_b)^2$. The Wilson coefficients of the dimension-seven and eight operators are known only to LO-QCD.

\begin{figure}[t]
\includegraphics[width=0.5\textwidth]{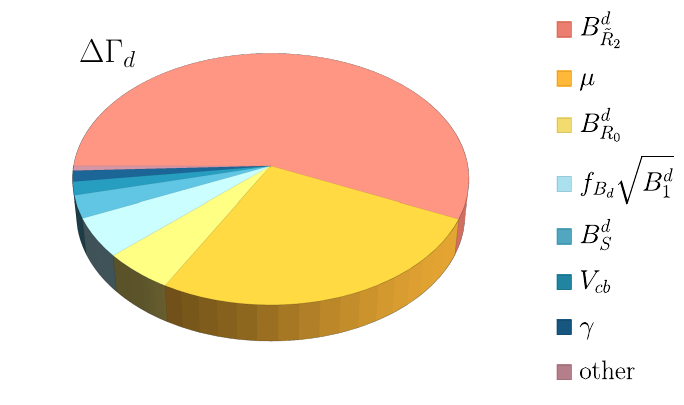}
\caption{Composition of the theoretical error in the SM prediction of 
$\Delta \Gamma_d$.}
\label{fig:Delta_Gamma_d}      
\end{figure}

\begin{figure}[t]
\includegraphics[width=0.5\textwidth]{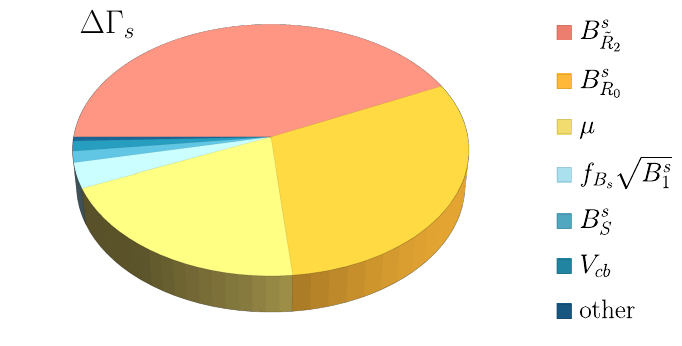}
\caption{ Composition of the theoretical error in the SM prediction of 
$\Delta \Gamma_s$.}
\label{fig:Delta_Gamma_s}      
\end{figure}

We update the SM prediction for the decay rate difference from Ref.~\cite{Lenz:2019lvd} with new input for the CKM matrix elements and new averages (including correlations) of lattice QCD and HQET sum rule results for the matrix elements of the four-quark operators \cite{Greljo:2022jac}, to obtain
\begin{eqnarray}
\Delta \Gamma_d & = & 
(2.7 \pm 0.4) \cdot 10^{-3} \, \mbox{ps}^{-1},
\label{eq:DeltaGamma-d-SM}
\\
\Delta \Gamma_s & = & 
(9.1 \pm 1.5) \cdot 10^{-2}\, \mbox{ps}^{-1}.
\label{eq:DeltaGamma-s-SM}
\end{eqnarray}
The composition of the theory  uncertainties are depicted in Fig.~\ref{fig:Delta_Gamma_d} and Fig.~\ref{fig:Delta_Gamma_s}, with the largest uncertainties stemming from the matrix elements of the genuine dimension-seven operators.

\begin{figure}[t]
    \centering
    \includegraphics[width=0.5\textwidth]{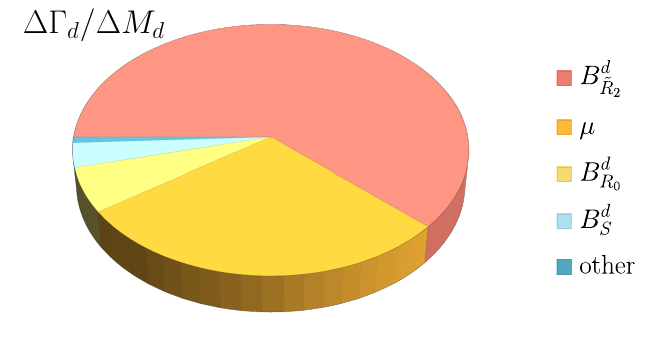} \\
     \caption{ Composition of the relative theory uncertainties  in the SM prediction of 
$\Delta \Gamma_d/\Delta M_d$}
 \label{fig:pie-charts-DeltaGamma-to-DeltaM-d}
\end{figure}
\begin{figure}[t]
    \includegraphics[width=0.5\textwidth]{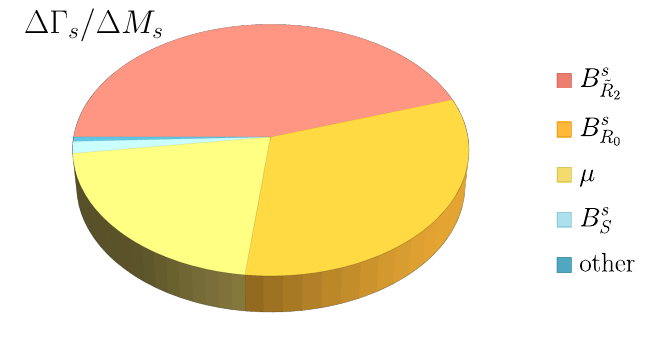}
    \caption{ Composition of the relative theory uncertainties in the SM prediction of $\Delta \Gamma_s/\Delta M_s$}
    \label{fig:pie-charts-DeltaGamma-to-DeltaM-s}
\end{figure}

In our analysis they result in a relative uncertainty of around $14 \%$, while the numerical analysis in Ref.~\cite{Gerlach:2022hoj} obtains for the same contribution a considerably larger uncertainty of $18 \% - 20 \% $.
Due to its small numerical effect we have not included the NNLO-QCD corrections $\tilde{\Gamma}_6^{(2)}$ \cite{Gerlach:2021xtb,Gerlach:2022wgb,Gerlach:2022hoj}, but we nevertheless would like to emphasize the importance of knowing the size of the NNLO-QCD effect.
In Ref.~\cite{Gerlach:2022hoj}, it is further argued to determine the decay rate difference from the SM value of the ratio $\Delta \Gamma_s/\Delta M_s$ and the experimental value of the mass difference via
\begin{eqnarray}
\Delta \Gamma_q & = & 
\left( \frac{\Delta \Gamma_q }{\Delta M_q }
\right)^{\rm HQE} 
\!\!\!\!\!\!
\cdot
\,\,\,
\Delta M_q^{\rm Exp.} \, ,
\label{eq:trick17}
\end{eqnarray}
since in the theory ratio the dependence on the decay constant and the dominant CKM dependence cancel. We note, however, that currently the theory uncertainty in $\Delta \Gamma_q$ is completely dominated by the dimension-seven contributions, and the theory uncertainty of the ratios is with $16.0\%$ only marginally smaller than the theory uncertainty of the decay rate difference determined via Eq.~(\ref{eq:def_DeltaGamma}), which tuned out to be $16.3\%$ .
Moreover, the use of Eq.~(\ref{eq:trick17}) relies on the assumption that the experimental value of the mass difference is given exactly by the SM prediction, hence we prefer  the direct determination with the help of Eq.~(\ref{eq:def_DeltaGamma}), resulting in the values presented in Eqs.~\eqref{eq:DeltaGamma-d-SM}, \eqref{eq:DeltaGamma-s-SM}.

For the ratio $\Delta \Gamma_q/\Delta M_q$ we obtain
\begin{eqnarray}
\left( \frac{\Delta \Gamma_d }{\Delta M_d }
\right)^{\rm HQE} & = & 
(50.5 \pm 6.8) \cdot 10^{-4} \, \, ,
\\
\left( \frac{\Delta \Gamma_s }{\Delta M_s }
\right)^{\rm HQE} & = & 
(49.9 \pm 7.9) \cdot 10^{-4} \, 
 \, .
\end{eqnarray}
The compositions of the relative uncertainties squared in the above predictions are depicted in Fig.~\ref{fig:pie-charts-DeltaGamma-to-DeltaM-d} and~\ref{fig:pie-charts-DeltaGamma-to-DeltaM-s}.

Using Eq.~\eqref{eq:trick17} and experimental value of the mass difference $\Delta M_s$ in Eq.~\eqref{eq:DeltaM-exp}, we find for the decay rate difference $\Delta \Gamma_s  =  (8.8 \pm 1.4) \cdot 10^{-2}\, \mbox{ps}^{-1} $, which agrees within uncertainties with the value of $\Delta \Gamma_s  =  (7.6 \pm 1.7) \cdot 10^{-2}\, \mbox{ps}^{-1} $ quoted in Ref.~\cite{Gerlach:2022hoj}. 
In our analysis, only the $\overline{MS}$-scheme is used, while Ref.~\cite{Gerlach:2022hoj}
averages over different renormalisation scales, which
leads to a further, small reduction of their central value.

\begin{figure}[t]
\includegraphics[width=0.5\textwidth]{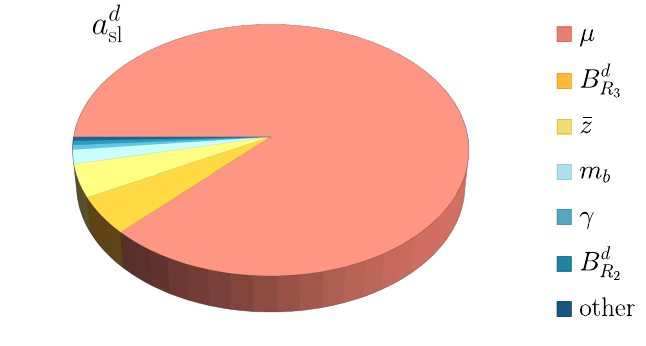}
\caption{Composition of the theoretical error in the SM prediction of 
$a_{sl}^d$.}
\label{fig:a_sl_d}      
\end{figure}

\begin{figure}[t]
\includegraphics[width=0.5\textwidth]{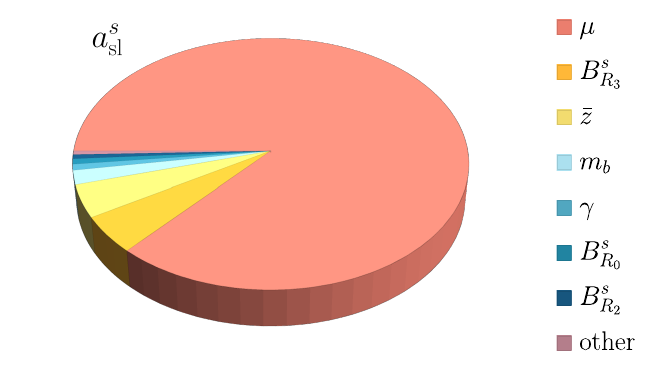}
\caption{ Composition of the theoretical error in the SM prediction of 
$a_{sl}^s$.}
\label{fig:a_sl_s}      
\end{figure}

Finally, we also update the SM predictions for the 
semi-leptonic CP asymmetries 
\begin{eqnarray}
a_{sl}^d & = & 
 -(5.1 \pm 0.5) \cdot 10^{-4} \, ,
\label{eq:asl-SM-d}
\\
a_{sl}^s
 & = & 
+ (2.2\pm 0.2) \cdot 10^{-5} \,  ,
\label{eq:asl-SM-s}
\end{eqnarray}
with the compositions of the relative theory uncertainties squared depicted in Fig.~\ref{fig:a_sl_d} and Fig.~\ref{fig:a_sl_s}. The largest uncertainties originate from the renormalisation scale dependence followed by the charm quark mass dependence. 
The  NNLO-QCD results of \cite{Gerlach:2021xtb,Gerlach:2022wgb,Gerlach:2022hoj} 
are only known to leading  power in $(m_c/m_b)^2$, which cancel exactly in the GIM suppressed expression for the semi-leptonic asymmetries. To determine NNLO-QCD corrections to the semi-leptonic CP asymmetries  higher powers in $(m_c/m_b)^2$ will have to be determined.

\subsection{Experiment}
\label{sec:mix_exp}
\subsubsection{$B^0$ mixing: measurement of $\Delta \Gamma_d$ and $\Delta M_d$}
\label{sec:mix_exp_Bd}
A large number of time-dependent analyses of the $B^0 - \bar{B}^0$ oscillation have been performed in the past 20 years by the ALEPH, DELPHI, L3, OPAL, CDF, D\O\, \linebreak BABAR, Belle and LHCb collaborations, they are summarized in Ref.~\cite{HeavyFlavorAveragingGroup:2022wzx}. HFLAV performs a combination of all relevant measurements, including the identified correlations and systematics. The combined $B^0$ mixing frequency is found to be~\cite{HeavyFlavorAveragingGroup:2022wzx} 
\begin{equation}
\Delta M_d = 0.5065 \pm 0.0016 \pm 0.0011\, \mbox{ps}^{-1}.
\end{equation}

The situation for $\Delta \Gamma_d$ is less well known, $\Delta \Gamma_d / \Gamma_d$ is too small and the knowledge on the time integrated mixing probability is too imprecise to provide useful sensitivity on $\Delta \Gamma_d / \Gamma_d$. Direct time-dependent studies provide much stronger constraints: the most stringent individual bounds come from Belle and have a precision below 2\%~\cite{hflav166}. 

Measurements of LHCb~\cite{LHCb:2014qsd}, ATLAS~\cite{ATLAS-168} and \linebreak CMS~\cite{CMS:2017ygm} compare measurements of the lifetime for $B^0 \to J /\psi K^{*0}$ and $B^0 \to J /\psi K^0_S$ decays, which provide input to the HFLAV average \cite{HeavyFlavorAveragingGroup:2022wzx}
\begin{eqnarray}
\frac{\Delta \Gamma_d}{\Gamma_d}  = 0.001 \pm 0.010\,.
\end{eqnarray}
Several methods exist to improve the experimental knowledge of $\Delta \Gamma_d / \Gamma_d$, a summary can be found in Ref.~\cite{Gershon:2010wx}, where this observable is, due to its extremely small SM value, also advocated as a null test of the Standard Model. 

\subsubsection{$B^0_s$ mixing: measurement of $\Delta \Gamma_s$ and $\Delta M_s$}
\label{sec:Bs-Mix_Exp}

\begin{figure}
 \includegraphics[width=0.5\textwidth]{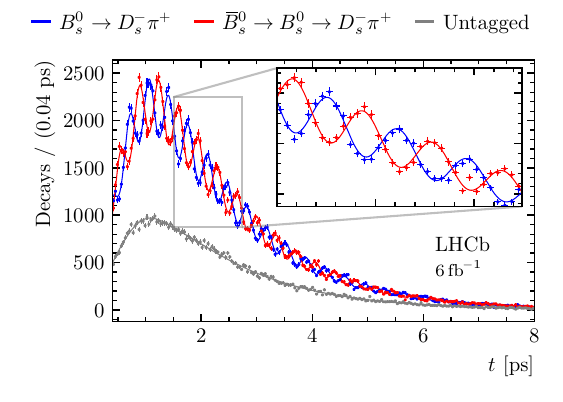}
\caption{Distribution of the decay time of the $B_s^0 \to  D_s^- \pi^+$ signal decays. The three components, unmixed, mixed and untagged, are shown in blue, red and gray, respectively. The insert corresponds to a zoom of the region delineated in grey.
Figure reproduced from  Ref.~\cite{LHCb-201}.}
\label{fig:bs-dms}
\end{figure}

Oscillations of neutral $B^0_s$ mesons were first observed by the CDF collaboration in 2006~\cite{CDF-197} based on samples of flavour-tagged hadronic and semileptonic $B_s^0$ decays in flavour-specific final states. The LHCb collaboration has published the most precise results on $B^0_s$ oscillations using fully reconstructed $B_s^0 \to  D_s^- \pi^+$ and $B_s^0 \to D_s^- \pi^+ \pi^- \pi^+$ decays~\cite{LHCb-198,LHCb-199,LHCb-200,LHCb-201}. Fig.~\ref{fig:bs-dms} shows the distribution of the decay time of the $B_s^0 \to  D_s^- \pi^+$ signal decays, where the mixed and unmixed contributions have been identified. 

An average of all the published CDF, LHCb and CMS results~\cite{HeavyFlavorAveragingGroup:2022wzx} yields
\begin{equation}
    \Delta M_s = 17.765 \pm 0.004 \pm 0.004 \, \mbox{ps}^{-1}.
\end{equation}
This experimental value is consistent, but much more precise than the current SM value for $\Delta M_s$. 

The best sensitivity to decay rate difference $\Delta \Gamma_s$ is currently achieved by the recent time-dependent measurements of $B_s^0 \to (cc) K^+ K^- $ decay rates performed at CDF, D\O\, ATLAS, CMS and LHCb, where the CP-even and CP-odd amplitudes are statistically separated through a full angular analysis. Information on $\Delta \Gamma_s$ is also obtained from the study of the proper time distribution of untagged samples of flavour-specific $B_s^0$ decays and from the effective lifetimes measured in pure CP modes.
These experimental results are averaged~\cite{HeavyFlavorAveragingGroup:2022wzx} to 
\begin{eqnarray}
\Gamma_s
&=& 
0.6573 \pm 0.0023 \, \mbox{ps}^{-1},
\\
\Delta \Gamma_s 
& = & 
+ 0.083 \pm 0.005 \, \mbox{ps}^{-1},
\\
\frac{\Delta \Gamma_s}{\Gamma_s}  
& = & 
+ 0.126 \pm 0.007,
\end{eqnarray}
which are generally in good agreement with the SM predictions. However, some tensions exist in the individual measurements: the angular analyses of $B^0_s \to J/\psi \phi$ performed by ATLAS~\cite{ATLAS:2016pno,ATLAS:2020lbz}, CMS~\cite{CMS:2015asi,CMS:2020efq} and LHCb~\cite{LHCb:2014iah,LHCb-183} show some tensions at the level of approximately $3\sigma$, driven by the time and angular distribution parameters. Also, the ATLAS measurements~\cite{ATLAS:2020lbz} of $\Gamma_s$ show a tension of about 3$\sigma$ with the average, see Fig.~\ref{fig:Bs}. 
These tensions remain to be understood. 

\subsection{Phenomenology}
\label{sec:mix_pheno}
\subsubsection{Mass differences}
\begin{figure}[t]
\includegraphics[width=0.48\textwidth]{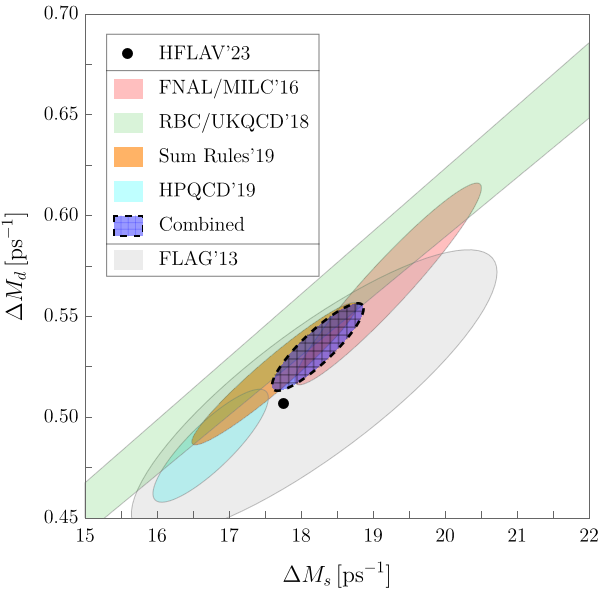}
\caption{ Comparison of different SM predictions for the mass differences $\Delta M_d$ and $\Delta M_s$ by using different determinations of the dimension-six matrix elements, as indicated in the legend, see also the text for more detail, with the corresponding HFLAV average denoted by the black dot 
(the latter has tiny uncertainties, practically invisible within the scale of the plot).}
\label{fig:DeltaMvsExp}      
\end{figure}

Comparing the values in Eqs.~\eqref{eq:DeltaM-exp}, \eqref{eg:DeltaM-d-SM}, \eqref{eg:DeltaM-s-SM}, one can see that 
experiment and theory agree well for the mass differences $\Delta M_d$ and
$\Delta M_s$
\begin{eqnarray}
\frac{\Delta M_d^{\rm SM}}{\Delta M_d^{\rm Exp.}} 
& = & 
1.056 \pm 0.042 \, ,
\hspace{0.5cm}
\frac{\delta \Delta M_d^{\rm SM}}{\delta \Delta M_d^{\rm Exp.}}
\simeq 11 \, ,
\label{eq:DMd_SM_Exp}
\\
\frac{\Delta M_s^{\rm SM}}{\Delta M_s^{\rm Exp.}} 
& = & 
1.026 \pm 0.036 \, ,
\hspace{0.5cm}
\frac{\delta \Delta M_s^{\rm SM}}{\delta \Delta M_s^{\rm Exp.}}
\simeq 105 \, ,
\label{eq:DMs_SM_Exp}
\end{eqnarray}
while the theoretical errors are much larger than the experimental ones.
Note that the above SM predictions correspond to the averaged Lattice QCD and HQET sum rule values of the dimension-six four-quark operator matrix elements. However, from the Lattice QCD side, for $\langle \tilde{Q}_{6,1} \rangle$ there is a slight tension between the large values obtained by FNAL-MILC~\cite{FermilabLattice:2016ipl}, leading to larger values of $\Delta M_{d,s}$ and the lower values obtained by the HPQCD-collaboration \cite{Dowdall:2019bea}. 
The HQET sum rules values~\cite{Kirk:2017juj,King:2019lal} lie in between 
the HPQCD and FNAL/MILC values.
Here an independent, further lattice evaluation would be highly desirable.
RBC/UKQCD has so far only investigated the $SU(3)_F$ breaking ratio $\Delta M_s / \Delta M_d$ \cite{Boyle:2018knm}, while a calculation of the individual bag parameters for the $B_d$ and $B_s$ system is currently in progress by JLQCD and RBC/UKQCD \cite{Boyle:2021kqn}.
In Fig.~\ref{fig:DeltaMvsExp}, we provide an updated version of the $\Delta M_d$ vs $\Delta M_s$ plot in Fig.~1 of Ref.~\cite{DiLuzio:2019jyq}, using the most recent values for the CKM parameters and the new averages of the non-perturbatvie parameters obtained in Ref.~\cite{Greljo:2022jac}.

In the SM, the mass differences are governed by a second order weak process and thus they are very suppressed. 
Hence, BSM models might give comparably large contributions, in particular, those ones which allow mixing of neutral $B$-mesons at tree-level.
In the light of current $B$-anomalies the $Z'$-boson and leptoquark models are considered as popular extensions of the SM that can accommodate these anomalies. Among them, the $Z'$-models, which explain the $B$-anomalies with a tree-level $bs-Z'$ coupling, create also $B_s$-mixing at tree-level. 
In 2017, the SM prediction for $\Delta M_s$ was governed by the large FNAL/MILC value and the $Z'$ contribution was further enhancing the prediction for the mass difference $\Delta M_s$. 
Thus, at that time a $Z'$-model explanation of the $B$-anomalies was largely excluded by $\Delta M_s$~\cite{DiLuzio:2017fdq}.
Redoing this analysis after the sum rule values and HQPCD values for $\Delta M_s$ became available, which reduced the SM expectation considerably, a $Z'$-model became viable again~\cite{DiLuzio:2019jyq}. This example shows nicely the importance of precise SM predictions for the mixing parameters and its crucial implications for potential BSM models, see also e.g.~Ref~\cite{DeBruyn:2022zhw}.
\subsubsection{Decay rate differences}
The decay rate difference has so far only been measured for the $B_s$ system.
Its first measurement by the LHCb collaboration in 2012
\cite{Delta Gamma} presented a nice test of the theoretical tools to describe the mixing quantities \cite{Lenz:2012mb}.
Comparing values in Eqs.~\eqref{eq:DeltaGamma-exp} and \eqref{eq:DeltaGamma-s-SM}, one can see that experiment and theory agree very well for $\Delta \Gamma_s$,
\begin{eqnarray}
 \frac{\Delta \Gamma_s^{\rm SM}}{\Delta \Gamma_s^{\rm Exp.}} & = & 1.10 \pm 0.19 \, ,
 \hspace{0.7cm}
 \frac{\delta \Delta \Gamma_s^{\rm SM}}{\delta \Delta \Gamma_s^{\rm Exp.}}
 \simeq 3 \, ,
 \label{eq:DGammas_SM_Exp}
 \end{eqnarray}
with experiment being roughly a factor of three more precise than experiment.
Besides being a further testing ground for our theoretical tools, decay rate differences are also sensitive to BSM effects in tree-level decays and light new physics, see e.g. Refs.~\cite{Lenz:2019lvd,Brod:2014bfa,Bobeth:2014rda,Crivellin:2023saq,Jager:2017gal,Jager:2019bgk,Bobeth:2011st,Bordone:2023ybl}.

Due to its smallness in the SM, future measurements of $\Delta \Gamma_d$ offer interesting possibilities as Null tests \cite{Gershon:2010wx}.

\subsubsection{Flavour specific CP asymmetries}
In the SM, flavour-specfic CP asymmetries are both CKM and GIM suppressed and thus expected to be tiny.
Currently, the experimental uncertainty of $a_{fs}^s$, see Eq.~\eqref{eq:asl-exp}, is roughly 140 times as large as the SM central value in Eq.~\eqref{eq:asl-SM-s}, 
while the experimental uncertainty in $a_{fs}^d$ is roughly three times as large as the corresponding SM central value,
see Eqs.~\eqref{eq:asl-exp} and \eqref{eq:asl-SM-d}. 
This leaves plenty of space for future discoveries of BSM effects. A particular sensitivity of the flavour-specific CP
asymmetries to new effects in non-leptonic
tree-level decays was worked out e.g. in
Refs.~\cite{Gershon:2021pnc,Lenz:2022pgw,Fleischer:2016dqd}.

\subsection{Outlook to the Charm system}
\label{sec:mix_charm}
The success of the HQE to describe also charmed hadron lifetimes suggests to apply the same framework for charm mixing quantities, see e.g. the review \cite{Lenz:2020awd}.
Charm mixing suffers, however, from an extreme GIM~\cite{Glashow:1970gm} suppression.
Analyses of exclusive decays \cite{Falk:2001hx,Falk:2004wg}, using simplified assumptions, like  taking only phase space effects into account but no dynamical QCD contributions, lead to a range of values for the mixing parameters that is consistent with the experimental data~\cite{HeavyFlavorAveragingGroup:2022wzx}, whereas studies computed within the HQE yield extremely suppressed results, see e.g.~Ref.~\cite{Bobrowski:2010xg}.
Despite recent progress made in assessing the uncertainty of the HQE prediction~\cite{Lenz:2020efu}, it is still unclear how to deal with such a strong GIM suppression from a theoretical point of view.

Neglecting the SM contribution to $D$-mixing and attributing the experimental values to BSM effects 
can nethertheless provide interesting and important bounds on extensions of the SM, see e.g. Refs.~\cite{Chala:2019fdb,Dery:2019ysp,Bause:2022jes}.

\section{Conclusion and outlook}

Observables in $B$-mixing and $b$-hadron lifetimes are important tools to test and increase our understanding of quantum chromodynamics and for indirect searches for BSM effects. The historical evolution as well as the current status of these observables has been reviewed both from an experimental and theoretical point of view. 
In particular, we presented new updates for the SM predictions of the mixing observables $\Delta M_q$, $\Delta \Gamma_q$ and $a_{fs}^q$, based on most recent values for the CKM matrix elements and new averages of the non-perturbative parameters.
All in all, we find an excellent agreement of the  experimental data for lifetimes and mixing with our theoretical predictions. 
Therefore, we conclude, that our theory tools work very well and to a good approximation the SM gives the dominant contribution to most of the observables, while still leaving room for possible BSM effects. 

On the theory side, the predictions can be systematically improved and the uncertainties can be further reduced both by perturbative and non-perturbative calculations. 
In particular, we would like to mention  non-perturbative determinations of the matrix elements of dimension-seven operators for mixing, lattice QCD determination of matrix elements of four-quark operators contributing to lifetimes, and NNLO-QCD corrections to the free-quark decay, to the four-quark operator contributions like Pauli-interference, and NNLO corrections to $\Gamma_{12}^q$ including higher orders in the $m_c^2/m_b^2$ expansion, crucial for predicting the flavour specific CP asymmetries. 
Moreover, in the light of the large Darwin operator contribution to lifetime ratios, a first determination of the dimension-seven two-quark operator contribution would be highly desirable.

On the experimental side, some obervables like the mass difference $\Delta M_s$ or lifetimes of heavy hadrons exceed by far the theoretical precision, while others, like the decay rate difference $\Delta \Gamma_s$ or the lifetime ratios of baryons, have similar uncertainties as the corresponding theory determinations. In view of the  expected theoretical improvements further more precise data are desirable.
Some experimental baryon lifetime ratios, like $\tau(\Omega_b) / \tau (B_d)$ or $\tau(\Xi_b^0) / \tau (\Xi_b^-)$ still have larger (or comparable) uncertainties than the HQE predictions. 
Finally, there are also observables like the decay rate difference $\Delta \Gamma_d$ or the flavour-specific CP-asymmetries $a_{fs}^q$, where we are still lacking a non-zero measurement and have thus  plenty of space for BSM effects. 
Here any improvement of the current bounds will be very insightful.

Theory can also benefit from experimental measurements outside the field of lifetimes and mixing.
A combined effort from proton colliders (e.g. LHCb) and electron colliders (BES-III and Belle-II) in studying inclusive semi-leptonic decays of $B_d$-, $B^+$- and $B_s$-mesons could help to gain insight in $SU(3)_F$ breaking and the HQE expansion of the lifetimes of the $B_s$-meson and settle the current discrepancy originating in the largely unknown size of the matrix element of the Darwin operator.

Lifetimes of $b$-hadrons and mixing of neutral $B$-mesons will continue to be an exciting research field in the coming years and maybe some unexpected discoveries will show up in the interplay of color and flavor.
 
\begin{acknowledgements}
We would like to thank Quentin F\"uh\-ring  and Eleftheria Malami for critical reading of the manu\-script. J.\,A.\ acknowledges support from the Heisenberg program of the Deutsche Forschungsgemeinschaft (DFG), GZ: AL 1639/5-1 as well as within the Collaborative Research Center SFB1491 and from the German Federal Ministry of Education and Research (BMBF, grant no. 05H21PECL1) within ErUM-FSP T04. F.\,B.\, acknowledges support from the Emmy-Noether program of the DFG (BE 6075/1-1) and BMBF (grant no. 05H21PDKB). A.\,L.\ and A.\,R.\ acknowledge support from the DFG, under grant 396 021 762 - TRR 257 “Particle Physics Phenomenology after the Higgs Discovery” and by the BMBF project Theoretische Methoden für LHCb und Belle II (Förder-kennzeichen 05H21PSCLA/ErUM-FSP T04).
\end{acknowledgements}

\noindent
Data Availability Statement: No Data associated
in the manuscript.

\end{document}